\documentclass[journal]{IEEEtran}

\usepackage{courier}
\usepackage[usenames,dvipsnames]{color} 
\usepackage{color,url,xspace}
\usepackage{amssymb,amsmath}
\usepackage{graphicx}
\usepackage{colordvi}
\usepackage{epsfig}
\usepackage{endnotes}
\usepackage{verbatim}
\usepackage{subfigure}
\usepackage{textcomp}
\usepackage{subfig}
\usepackage{setspace}
\usepackage{xcolor}
\usepackage{amsmath}
\usepackage{cite}
\usepackage{times}
\usepackage{booktabs, multirow}
\usepackage{amsthm}
\usepackage{nccmath}
\usepackage{algorithmicx,algpseudocode}
\usepackage{algpseudocode,algorithm,algorithmicx}
\usepackage{caption}
\usepackage{mathtools}
\usepackage{mathtools, cuted}
\usepackage{lipsum, color}
\usepackage{stfloats}
\usepackage{supertabular}

\algrenewcommand\algorithmicrequire{\textbf{Precondition:}}
\algrenewcommand\algorithmicensure{\textbf{Postcondition:}}

\usepackage{color}
\usepackage{soul}

\begin{document}
\title{Wireless Video Caching and Dynamic Streaming under Differentiated Quality Requirements}
\author{
	\IEEEauthorblockN{
		Minseok Choi,~\IEEEmembership{Member,~IEEE,}
		Joongheon Kim,~\IEEEmembership{Senior Member,~IEEE,}
		and~Jaekyun Moon,~\IEEEmembership{Fellow,~IEEE}
		\thanks{Manuscript received December 1, 2018; revised April 9, 2018; accepted April 19, 2018.}
		\thanks{
			This work is in part supported by the National Research Foundation of
			Korea under Grant No. 2016R1A2B4011298, and in part supported by the National Research Foundation of Korea under Grant: 2016R1C1B1015406. J. Kim is a corresponding author of this paper.}
		\thanks{M. Choi and J. Moon are with the Department of Electrical Engineering, Korea Advanced Institute of Science and Technology (KAIST), Daejeon, Korea e-mails: ejaqmf@kaist.ac.kr, jmoon@kaist.edu.}
		\thanks{J. Kim is with the School of Computer Science and Engineering, Chung-Ang University, Seoul, Korea e-mail: joongheon@cau.ac.kr.}
	}
}

\markboth{IEEE Journal on Selected Areas in Communications}%
{}

\maketitle

\begin{abstract}

This paper considers one-hop device-to-device (D2D)-assisted wireless caching networks that cache video files of varying quality levels, with the assumption that the base station can control the video quality but cache-enabled devices cannot. 
Two problems arise in such a caching network: 
\textit{file placement problem} and \textit{node association problem}. 
This paper suggests a method to cache videos of different qualities, and thus of varying file sizes, by maximizing the sum of video quality measures that users can enjoy. 
There exists an interesting trade-off between video quality and video diversity, i.e., the ability to provision diverse video files.
By caching high-quality files, the cache-enabled devices can provide high-quality video, but cannot cache a variety of files. 
Conversely, when the device caches various files, it cannot provide a good quality for file-requesting users. 
In addition, when multiple devices cache the same file but their qualities are different, advanced node association is required for file delivery. 
This paper proposes a node association algorithm that maximizes time-averaged video quality for multiple users under a playback delay constraint. 
In this algorithm, we also consider \textit{request collision}, the situation where several users request files from the same device at the same time, and we propose two ways to cope with the collision: scheduling of one user and non-orthogonal multiple access. 
Simulation results verify that the proposed caching method and the node association algorithm work reliably.

\end{abstract}

\begin{IEEEkeywords}
Wireless caching network, D2D communication, Video streaming, Caching policy, Node association
\end{IEEEkeywords}

\IEEEpeerreviewmaketitle

\section{Introduction}
\IEEEPARstart{E}{xceedingly} large amounts of data traffic generated by rapidly growing wireless mobile devices in recent years have created formidable challenges for  wireless communication.
Within just a few years, it is expected that tens of exabytes of global data traffic be handled on daily basis with on-demand video streaming services accounting for about 70\% of them~\cite{cisco}.   

On-demand video streaming is characterized by a relatively small number of popular contents being requested at ultra high rates; as such playback delay is often the more important measure of goodness to the user than other typical performance metrics like video quality \cite{youtube}.
In this regard, the wireless caching technology as discussed in \cite{caching,caching_5G}, wherein the base station (BS) pushes popular contents for off-load time to cache-enabled nodes with limited storage spaces so that these nodes provide popular contents directly to nearby mobile users, is advantageous for video streaming services. 
By caching popular files on cache-enabled nodes, there is no need to repeatedly receive files from the BS every time users request.

Caching popular contents on the finite storage of the helper node near mobile users, which acts like a small BS, has been proposed to reduce latency in file transmission \cite{TIT2013shanmugam}.
Further, a device-to-device (D2D)-assisted caching network has been studied \cite{CM2013golrezaei,JSAC2016ji,TIT2016Ji,TWC2014golrezaei}, where mobile devices can store popular contents and directly respond to the file requests of neighboring users.
In the wireless caching network, there are two main issues: 1) \textit{file placement problem} - how to cache the popular contents at the caching nodes, e.g., caching helpers or cache-enabled devices, and 2) \textit{node association problem} - which caching node is optimal to deliver the requested file to the user for providing smooth video streaming services.

Video files can be encoded to multiple versions which differ in the quality level, e.g., peak-signal-to-noise-ratio (PSNR) or spatial resolution \cite{VideoEnc:ICME2002Hartanto,INFOCOM2014Poularakis}.
Since the file size of video varies by quality, it is also important in caching network to determine which file of what quality is stored in the caching node (\textit{file placement problem}) and 
what quality of video is requested from which caching node by the streaming user (\textit{node association problem}) \cite{VideoCaching:Network2017Argyriou}.

The goal of the file placement problem is to find the optimal caching policy according to popularity distribution of contents and network topology.
There have been some research efforts to find the optimal caching policy in stochastic wireless caching networks \cite{ICC2015blaszczyszyn,TWC2016chae,TC2016malak}, but contents with different quality levels were not considered.
Traditionally, caching strategies for videos with various qualities have been researched with radio access network (RAN) caches which enable transcoding or transrating of video files \cite{ABR:TOM2004Shen,ABR:TOM2013Zhang,ABR:TON2016Pedersen}.
However, deployments of the transcoder in mobile devices are inefficient, thus it is reasonable that 
only the video file of certain quality pushed by the BS for off-load time can be delivered by cache-enabled devices.

Due to the finite storage size of caching devices, there exists a trade-off between video quality and video diversity, i.e., if the device wants to cache the high-quality files, it cannot store many types of videos.
The authors of \cite{CachingDiffQual:WiOpt2016Jarray} consider caching files of different sizes, but they assume that the different-sized files account for the same unit of cache storage, thus it does not reflect the above trade-off.
Many researchers have proposed the static file placement policies under the consideration of differentiated quality requests for the same file, given probabilistic quality requests \cite{INFOCOM2014Poularakis,ITC2017Ye,CL2017Zhan} or minimum quality requirements \cite{INFOCOM2016Poularakis}.
In \cite{IFIP2016Araldo}, joint optimization of the static file placement and routing is proposed.
Further, the probabilistic caching policy for video files of various quality levels is presented in \cite{CL2016Wu} by using stochastic geometry, given the user preference for quality level.

The node association problem for video delivery in wireless caching networks has been also extensively researched.
In most of the research works that do not consider different quality levels for the same file, 
the file-requesting user is allowed to receive the content from the caching node under the strongest channel condition  \cite{TWC2016chae}, \cite{TWC2016yang}.
Node associations for video delivery in heterogeneous caching networks have been studied in \cite{TC2014Poularakis,TC2016zhang,TMC2017Jiang}.
Especially, dynamic video streaming allows each chunk, which consists of the whole video file and occupies a part of playback time, to have a different quality depending on time-varying network conditions \cite{DASH}. 
There are some research results addressing the transmission scheme which provides the video by dynamically selecting the quality level \cite{TOM2013Wang,TON2016kim} or the scheduling policy that maximizes a network utility function of time-averaged video quality in a network with caching helpers \cite{TC2015bethanabhotla}.
While the video delivery polices of \cite{TOM2013Wang,TON2016kim,TC2015bethanabhotla} are operated at the BS side, however, decisions of video delivery requests at user sides have been largely neglected.
This scenario is consistent with the practical real-world software implementation of dynamic adaptive streaming over HTTP (DASH) \cite{DASH}, in which users dynamically choose the most appropriate video quality.

In this paper, we consider the stochastic D2D-assisted caching network for dynamic video streaming services. 
For \textit{file placement problem}, each BS has all video files and is equipped with a quality controller, which controls the video quality. 
However, deploying the video quality controller in small mobile devices is not desirable; we assume in the present paper that the BS pushes the video files with certain quality levels and the cache-enabled devices can provide given video quality measures to mobile users who request the cached files.

For \textit{node association problem}, users dynamically request different quality levels for the same video and associate with one of the neighboring nodes caching the file of desired quality.
Assuming that delivered video chunks are waiting for playback in the user queue, the quality level of the next chunk should be chosen at the user side depending on user's channel condition and queue state to avoid playback delay, which is different from the assumption made in \cite{TC2015bethanabhotla}.
Depending on the desired quality, node association is updated for video delivery on chunk-by-chunk basis over the playtime.

In this paper, file placement and node association operate on different time scales, unlike \cite{JSAC2016gregori}, which jointly optimizes caching and transmission policies at the BS side. 
In general, the BS pushes popular contents to caching nodes for off-load time, and users request video files after file placement is completed. 
In addition, file popularity does not change as rapidly as dynamic changes of quality requests during video streaming, so file placement and node association are independently considered in this paper.

The main contributions of this paper can be summarized as follows:
\begin{itemize}
	
	\item This paper proposes the probabilistic caching policy for video files of varying quality levels by maximizing the successfully enjoyable video quality sum. 
	Since the streaming user dynamically requests the quality level of video, the expected quality of video which can be reliably delivered to the user, i.e., successfully enjoyable video quality summation, is a reasonable metric.
	We derive the closed-form caching probabilities for every video file of every quality level.
	The trade-off between video quality and video diversity is reflected in the proposed caching placement policy.
	
	\item This paper models the node association cases when video files of different quality levels are stored in cache-enabled devices.
	We specify the cases which require an advanced node association scheme to carefully choose the cache-enabled device for video delivery with desired quality.
	
	\item This paper proposes a node association algorithm for file-requesting users to choose the appropriate quality and to associate with the device which caches the requested file of desired video quality.
	The proposed algorithm maximizes the sum of the time-averaged quality measures of all users while avoiding playback delay in streaming communications.
	In this paper, playback delay is interpreted based on the user queue model, and the algorithm aims at avoiding playback delay by preventing queue emptiness.
	Simply, when there is no video chunk in the user queue, the user has to wait for the next chunk and video playback is inevitably delayed.
	Compared to pursuing only quality and only preventing playback latency, numerical results show that the proposed algorithm allows video chunks to be stacked in queue enough to maintain smooth video playback, while pursuing high video quality.
	
	\item We provide two ways to handle \textit{request collision}, which occurs when multiple users request video files simultaneously from the same cache-enabled device. 
	One is to schedule one of the file-requesting users for video delivery. 
	Another method is utilization of NOMA to serve all file-requesting users at the expense of data rate degradation.
	In the proposed algorithm, a scheduling scheme maximizes the time-average video quality for the given user while preventing playback delay.
\end{itemize}

The rest of the paper is organized as follows.
The D2D-assisted caching network model with different-quality video files is given in Section \ref{sec:NetworkModel}.
Caching policy for video files of various quality levels is proposed in Section \ref{sec:FilePlacement}.
Node association cases with different-quality video files and the node association algorithm are presented in Section \ref{sec:node_association}.
Simulation results are shown in Section \ref{sec:performance_evaluation} and Section \ref{sec:conclusion} concludes the paper.

\section{D2D Caching Network Model with Different-Quality Video Files}
\label{sec:NetworkModel}

\begin{figure} [h!]
	\centering
	\includegraphics[width=0.35\textwidth]{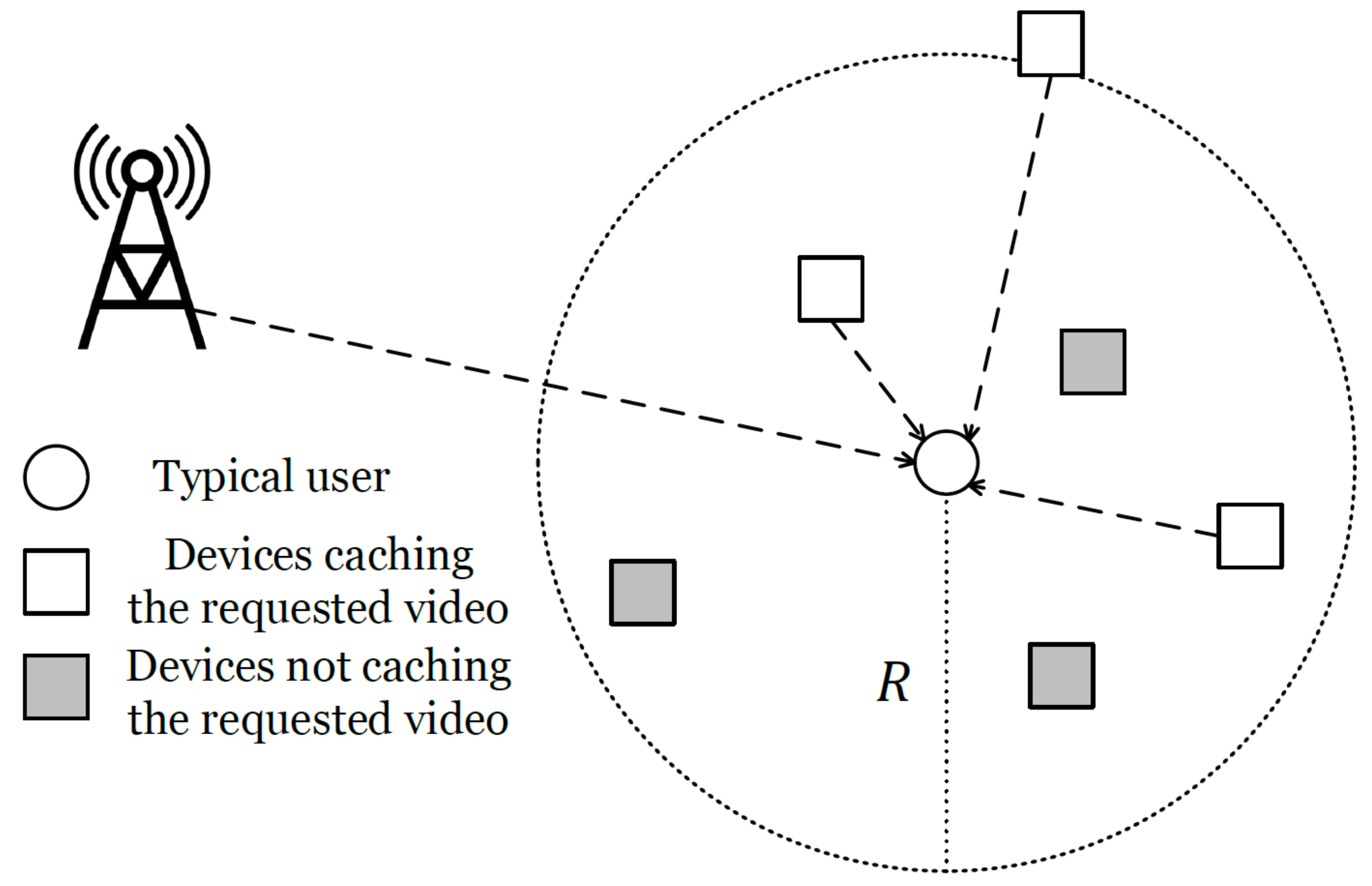}
	\caption{D2D Caching Network Model}
	\label{Fig:SystemModel}
\end{figure}

This paper considers a cellular model where some cache-enabled devices exist and $N$ users enjoy video streaming services. 
When certain user $n$ requests a particular video file, she searches through the device candidates that cache the requested file within a radius of $R$, as shown in Fig. \ref{Fig:SystemModel}.
User $n$ selects one of the candidates for file delivery.
If there is no device caching the requested file within the radius $R$ from user $n$, the BS can transmit the desired file via a cellular link. 
Since the caching devices are usually much closer to the file-requesting users than the BS, the users are assumed to prefer downloading the file from the caching devices rather than directly from the BS, due to transmission delay.
Therefore, direct transmission from the BS is not considered in this paper.

There is a file library $\mathcal{F}$ and each file $i\in \mathcal{F}$ has a popularity probability $f_i$, which follows the Zipf distribution \cite{TWC2014golrezaei}: $f_i= i^{-\gamma}/\sum_{j=1}^F j^{-\gamma}$ where $\gamma$ denotes the popularity distribution skewness.
Let $i_n$ be the index of the file requested by user $n$.
Assume that all files have $Q$ quality levels.
Suppose that there is no quality controller in cache-enabled devices, so devices can only transmit video files of the fixed quality which the BS pushes.
In this case, user $n$ can choose the quality level of the receiving video file; let $q_n$ denote the desired quality level of file $i_n$.
The file size varies with video quality, and let $M_q$ be the normalized file size of quality level $q$ for every video.
Each cache-enabled device has a limited storage size of $M$. 

The cache-enabled devices are modeled using the independent Poisson point processes (PPPs) with intensity $\lambda$.
This paper utilizes the probabilistic caching placement method \cite{ICC2015blaszczyszyn} for cache-enabled devices to cache file $i$ of quality $q$ with probability $p_{i,q}$.
Let $\lambda p_{i,q}$ be the intensity of the independent PPPs for the devices caching file $i$ of quality level $q$.
Suppose that the system does not allow any additional D2D link within the radius $R$ of the user who is already downloading the file from certain cache-enabled device.
By taking $R$ sufficiently large and/or exploiting orthogonal resources for each D2D coverage, the system can guarantee the negligible interference among multiple D2D links.
When an additional user requests a video file within the coverage, the user should download the file from the BS via the cellular link.

The Rayleigh fading channel is assumed for the communication links from the users to the cache-enabled devices. Denote the channel with $h = \sqrt{L} g$, where 
$L = 1/l^2$ controls slow fading with $l$ being the user-device distance and
$g$ represents the fast fading component having a complex Gaussian distribution, $g \sim CN(0,1)$.

The main research issues in the entire wireless caching network can be largely classified as follows:
\begin{itemize}
	\item \textit{File placement problem}: 
	When the BS pushes video files to cache-enabled devices for off-load time, the BS determines which file of which quality level is cached in each cache-enabled device.
	This paper chooses the probabilistic caching placement method \cite{ICC2015blaszczyszyn} for cache-enabled devices to cache file $i$ of quality $q$ with probability of $p_{i,q}$. 
	
	\item \textit{Node association problem}: 
	Each file-requesting user should find the candidate set of devices caching the requested video first. 
	Next, each user chooses one of the candidate devices for file delivery. 
	A careful choice of the device to be associated with is important to ensure good video quality and smooth playback without delay.
	
	\item \textit{Request collision}: 
	When multiple users request video files from the same device, we say \textit{request collision} occurs.
	In this instance, the device should determine how to serve those users.  
	One way is to deliver the requested file to only one user, expecting that each of the rest of the users finds another cache-enabled device to request video files. 
	The other method is NOMA, which serves multiple users in the same time/frequency/code simultaneously, but a transmission rate reduction is inevitable.
\end{itemize}

\section{Caching Policy for Different-Quality \\Video Files}
\label{sec:FilePlacement}

\subsection{Probabilistic Caching with Different Quality Video Files}

As mentioned earlier, the file placement problem in this paper is based on the probabilistic caching method \cite{ICC2015blaszczyszyn}, where the file is independently placed in devices according to the same distribution. 
Since we consider video files of 
different sizes, however, a certain modification of the probabilistic placement policy of \cite{ICC2015blaszczyszyn} is necessary.
As in \cite{ICC2015blaszczyszyn}, we also start with $M$ continuous memory intervals of unit length, and then place all files of all quality levels one by one to fill the $M$ unit-length intervals with every $p_{i,q}$.
The main difference from the approach of \cite{ICC2015blaszczyszyn} is that the file of quality level $q$ occupies a vertical size of $M_q$. 
Accordingly, we need to impose the following constraints:
\begin{align}
&\sum_{i=1}^F \sum_{q=1}^Q M_q p_{i,q} \leq M \label{eq:CachingConstraint}\\
&0 \leq p_{i,q} \leq 1, \forall i\in \mathcal{F},~\forall q \in \mathcal{Q}. \label{eq:CachingConstraint2}
\end{align}

The constraint (\ref{eq:CachingConstraint2}) is obvious, and the constraint (\ref{eq:CachingConstraint}) is necessary and sufficient for the existence of a random file placement policy requiring no more storage than $M$.
The sufficiency of (\ref{eq:CachingConstraint}) is proven by obtaining the caching policy requiring no more storage than $M$ in the following sections (see Table \ref{Table:CachingProb}).
The necessity of (\ref{eq:CachingConstraint}) can be also proven by establishing that the left-hand side of (\ref{eq:CachingConstraint}) is equal to the expected required memory size of the caching device, similar to Fact 1 in \cite{ICC2015blaszczyszyn}.
In addition, if the device caches file $i$ of quality $q_1$, then the same file of another quality level, say $q_2$, is better not to be cached in the device \cite{TMC2017Jiang}. 
However, it is not necessary to prevent caching copies of the same file with different qualities on a device for obtaining the caching policy.

\begin{figure} [h!]
	\centering
	\includegraphics[width=0.4\textwidth]{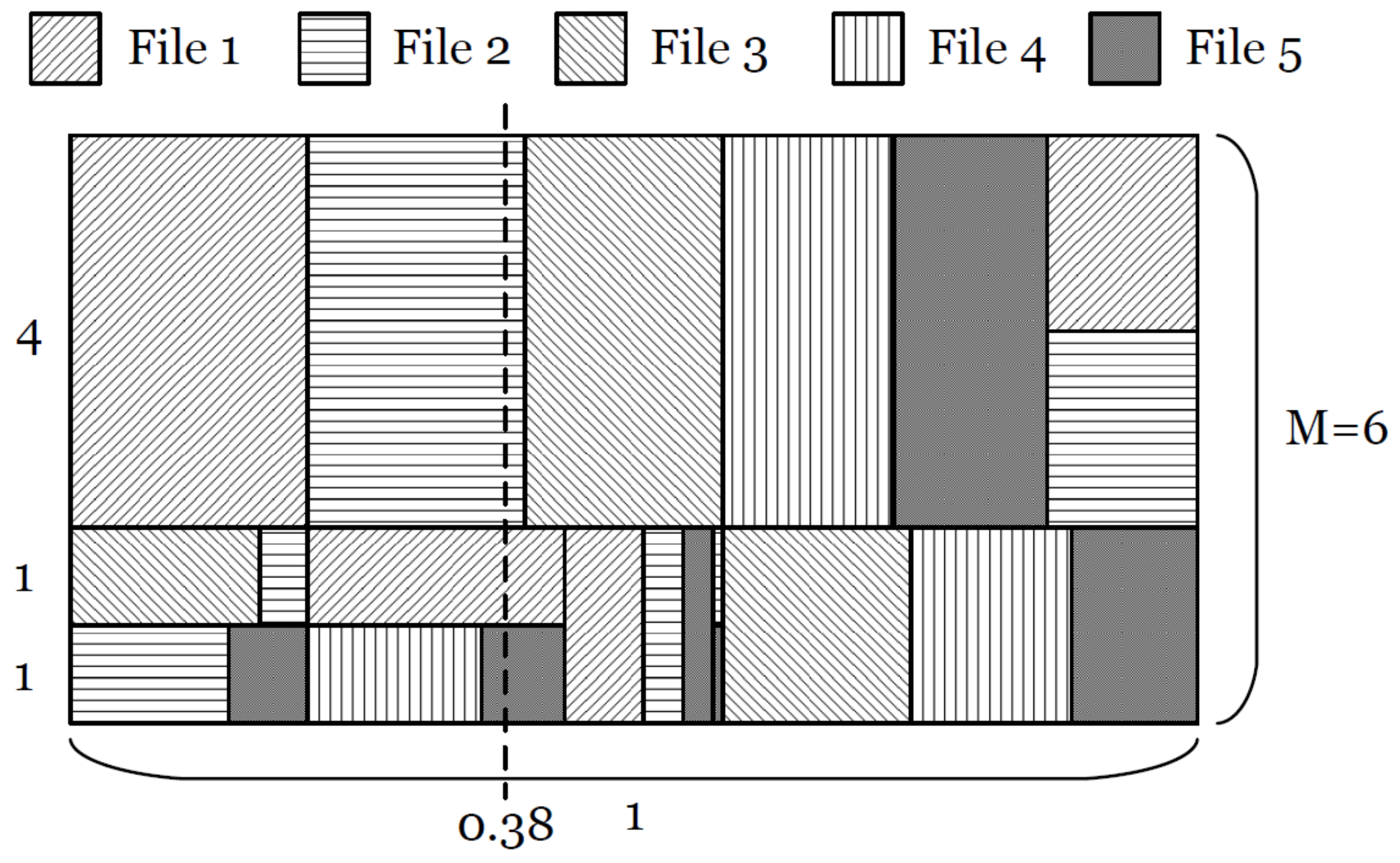}
	\caption{An example of probabilistic caching method with the files of different quality levels}
	\label{Fig:CachingExample}
\end{figure}

Fig. \ref{Fig:CachingExample} gives an example of the probabilistic caching method with files of different sizes where $F=5$, $M=6$, $Q=3$, $M_1 = 1$, $M_2 = 2$, and $M_3 = 4$.
This example satisfies the equality of (\ref{eq:CachingConstraint}).
As mentioned before, there are three kinds of blocks with vertical sizes $M_1=1$, $M_2=2$ and $M_3=4$ for each file type. 
After obtaining $p_{i,q}$, we have to build a $M \times 1$ rectangle consisting of $F \times Q$ rectangles with heights of $M_q$ and widths of $p_{i,q}$ for all $i \in \mathcal{F}$ and $q \in \mathcal{Q}$, as shown in Fig. \ref{Fig:CachingExample}.
Each element rectangle corresponds to the file of certain quality.
The cache-enabled device generates uniformly a random number within $[0,1]$, and draws a vertical line. 
Finally, the device stores files which the vertical line goes through.
In the example of Fig. \ref{Fig:CachingExample}, assuming we draw a vertical line at 0.38, the device stores File 2 of quality level 3, File 1 of quality level 1, and File 5 of quality level 1.

\textit{Remark}: Caching different-quality/different-size files would make storage inefficient in some system environments.
For example, consider the cases of $M_1 = 1$, $M_2=3$, $M_3 = 4$, and $M=6$.
In this case, a device cannot store two files of quality levels 2 and 3. 
The only possible combinations here are: caching one of quality 2 and three files of quality 1, and caching one of quality 3 and two files of quality 1. 
It is highly likely that the placements of $F \times Q$ rectangles with heights of $M_q$ and widths of $p_{i,q}$ do not fit perfectly in a $M\times 1$ rectangle in this scenario.
This situation indicates that storage is not being used efficiently. 
Therefore, the file sizes of different qualities and the maximum storage size of the device should be carefully considered for efficient caching.
However, this does not mean that the proposed constraints (\ref{eq:CachingConstraint}) and (\ref{eq:CachingConstraint2}) would not lead to random file placement policy of different qualities.

\subsection{Optimal File Placement Rule}

There still remains the important question: how to find the optimal $p_{i,q}$?
Since we assume that multiple D2D links do not interfere with one another, we can refer to the file placement rule in the noise-limited network \cite{TWC2016chae}.
The differences here are the constraint of the probabilistic caching method (\ref{eq:CachingConstraint}) and the optimization metric used. 
The method of \cite{TWC2016chae} maximizes the average file delivery success probability, but since we are concerned with the video quality, the successfully enjoyable video quality sum is chosen as the performance metric.
The successfully enjoyable video quality sum is defined as
\begin{equation}
\sum_{i=1}^F f_i \sum_{q=1}^Q \mathcal{P}(q) \cdot P\{ R_{i,q} \geq \rho_{i,q} \},
\label{eq:enjoyable_quality}
\end{equation}
where $\mathcal{P}(q)$ is the measure of the quality $q$, $R_{i,q}$ is the data rate of the user to download file $i$ of quality $q$, and $\rho_{i,q}$ is the threshold of the data rate for reliable transmission of file $i$ of quality $q$.
Denoting the Rayleigh fading channel from the user to the associated device for downloading file $i$ of quality $q$ by $h_{i,q}$, the data rate of the user for downloading file $i$ of quality $q$ in the noise-limited environment is given by 
\begin{equation}
R_{i,q} = \mathcal{B}\log_2 \Big( 1 + \frac{|h_{i,q}|^2}{\sigma^2} \Big),
\label{eq:data_rate1}
\end{equation}
where $\mathcal{B}$ is the bandwidth, 
assuming a unit transmit power and a normalized noise variance of $\sigma^2$. 
If the user desires the file $i$ of quality $q$ and there are multiple device candidates caching file $i$ of quality $q$, it is reasonable for the user to download the file from the device whose channel condition is the strongest among the candidates.

Since the channel power $|h_{i,q}|^2$ follows the chi-squared distribution, i.e., Nakagami-1 fading channel, according to \cite{TWC2016chae}, the reliable transmission probability can be obtained by
\begin{equation}
P\{ R_{i,q} \geq \rho_{i,q} \} = 1- \exp \Big\{ - \frac{\kappa p_{i,q}}{\sigma^2( 2^{\rho_{i,q}/\mathcal{B}}-1)}  \Big\},
\end{equation}
where $\kappa = \pi \lambda\Gamma(2)$.

Thus, we can formulate the optimization problem to find the optimal caching probabilities:
\begin{align}
&\{p^*_{i,q}\} = \underset{\{p_{i,q}\}}{\arg \max} \sum_{i=1}^F f_i \sum_{q=1}^Q \mathcal{P}(q) \cdot P\{ R_{i,q} \geq \rho_{i,q} \}   \\
&~~~~~~~= \underset{\{p_{i,q}\}}{\arg \min} \sum_{i=1}^F f_i \sum_{q=1}^Q \mathcal{P}(q) e^{ -C_{i,q} p_{i,q} }  \label{eq:PlacementProb} \\
&~~~~~~~\text{s.t. } \sum_{i=1}^F \sum_{q=1}^Q M_q p_{i,q} \leq M \label{eq:PlacementProb_const1} \\
&~~~~~~~~~~~0 \leq p_{i,q} \leq 1,~\forall i \in \mathcal{F},~\forall q \in \mathcal{Q} \label{eq:PlacementProb_const2} 
\end{align}
where $C_{i,q}=\frac{\kappa}{ \sigma^2(2^{\rho_{i,q}/\mathcal{B}}-1) }$.
Since 
$\frac{d^2}{d^2 p_{i,q}}\{ e^{-C_{i,q}p_{i,q}} \} \geq 0$ and the objective function in (\ref{eq:PlacementProb}) is the weighted function of $e^{-C_{i,q}p_{i,q}}$, the optimization problem (\ref{eq:PlacementProb}) is convex.

The Lagrangian function of the objective (\ref{eq:PlacementProb}) is given by
\begin{align}
&\mathcal{L}( \{p_{i,q}, \mu_{i,q}\}, \nu ) = \sum_{i=1}^F f_i \sum_{q=1}^Q \mathcal{P}(q) \cdot e^{-C_{i,q}p_{i,q}}\nonumber \\
&~~+~ \nu (\sum_{i=1}^F \sum_{q=1}^Q M_q p_{i,q} - M) + \sum_{i=1}^F \sum_{q=1}^Q \mu_{i,q} (p_{i,q}-1),
\label{eq:lagrangian}
\end{align}
and the derivative of (\ref{eq:lagrangian}) with respect to $p_{i,q}$, is 
\begin{equation}
\frac{\partial \mathcal{L} ( \{p_{i,q}, \mu_{i,q}\}, \nu ) }{\partial p_{i,q}} = -f_i \mathcal{P}(q) C_{i,q} e^{-C_{i,q} p_{i,q} } + \nu M_q + \mu_{i,q},
\end{equation}
where $\nu$ and $\mu_{i,q}$ are the nonnegative Lagrangian multipliers.
Then, the Karush-Kuhn-Tucker (KKT) conditions for the optimization problem (\ref{eq:PlacementProb}) are given by 
\begin{eqnarray}
\frac{\partial \mathcal{L} ( \{p_{i,q}, \mu_{i,q}\}, \nu) }{\partial p_{i,q}} &=& 0 \label{eq:kkt1} \\
\nu (\sum_{i=1}^F \sum_{q=1}^Q M_q p_{i,q} - M) &=& 0 \label{eq:kkt3} \\
\mu_{i,q} (p_{i,q}-1) &=& 0, \label{eq:kkt4}
\end{eqnarray}
(\ref{eq:PlacementProb_const1})-(\ref{eq:PlacementProb_const2}), and $\mu_{i,q}, \nu \geq 0$ for all $i \in \mathcal{F}$ and $q \in \mathcal{Q}$.

From (\ref{eq:kkt1}), we can obtain the optimal caching probabilities: 
\begin{equation}
p_{i,q}^* = \frac{1}{ C_{i,q}} \{ \ln(f_i \mathcal{P}(q) C_{i,q}) - \ln (\nu M_q + \mu_{i,q}) \}, ~\forall i,~\forall q.
\label{eq:optCachingProb}
\end{equation}
We can easily note from (\ref{eq:optCachingProb}) that the better the quality of the video file, the higher the probability of being stored in the device.
On the other hand, larger file size of the higher-quality video makes caching probability smaller and decreases video diversity.
Thus, the trade-off between video quality and video diversity is observed in (\ref{eq:optCachingProb}).
This trade-off depends on the constant value, $C_{i,q}$, and Lagrangian multipliers, $\nu$ and $\mu_{i,q}$.

The next step is to find the Lagrangian multipliers.
We can determine the intervals of the Lagrangian multipliers by categorizing the caching probability value into three cases.
First, when $p_{i,q}=0$, $\mu_{i,q}=0$ because of (\ref{eq:kkt4}).
To satisfy (\ref{eq:kkt1}), $v = \frac{f_i \mathcal{P}(q) C_{i,q}}{M_q}$, but it is impossible because $f_i$, $C_{i,q}$, and $M_q$ are different for $i$ and $q$.
We can set $\nu = \underset{i,q}{\max} \frac{f_i \mathcal{P}(q) C_{i,q}}{M_q}$ to guarantee $p_{i,q}^* \geq 0$. 
Therefore,
\begin{equation}
\nu \geq \frac{f_i \mathcal{P}(q) C_{i,q}}{M_q},~\text{if } p_{i,q}=0.
\label{eq:nu_interval1}
\end{equation}
When $0<p_{i,q}<1$, $\mu_{i,q}=0$ also, and $\nu = \frac{f_i \mathcal{P}(q) C_{i,q}}{M_q} e^{-\kappa p_{i,q} C_{i,q}}$ is obtained for (\ref{eq:kkt1}).
Therefore,
\begin{equation}
\frac{f_i \mathcal{P}(q) C_{i,q}}{M_q} e^{-C_{i,q}} < \nu < \frac{f_i \mathcal{P}(q) C_{i,q}}{M_q},~\text{if } 0<p_{i,q}<1.
\label{eq:nu_interval2}
\end{equation}
Finally, when $p_{i,q}=1$, if $\nu = 0$, $\mu_{i,q} = f_i \mathcal{P}(q) C_{i,q} e^{-C_{i,q}}$, otherwise, $\mu_{i,q}= f_i \mathcal{P}(q) C_{i,q} e^{-C_{i,q}} - \nu M_q$, according to (\ref{eq:optCachingProb}). 
To satisfy $\mu_{i,q} \geq 0$, 
\begin{equation}
\nu \leq \frac{f_i \mathcal{P}(q) C_{i,q}}{M_q} e^{-C_{i,q}},~\text{if } p_{i,q} = 1.
\label{eq:nu_interval3}
\end{equation}

From (\ref{eq:nu_interval1})-(\ref{eq:nu_interval3}), we can realize that $p_{i,q}$ and $\mu_{i,q}$ are functions of $\nu$, so we only need to find the optimal value of $\nu$ to obtain the optimal caching probabilities.
If $\nu \leq \min\{ \frac{f_i \mathcal{P}(q) C_{i,q} }{M_q}e^{-C_{i,q}},~\forall i \in \mathcal{F},~\forall q \in \mathcal{Q} \}$, $p_{i,q} = 1,~\forall i \in \mathcal{F},~\forall q \in \mathcal{Q}$.
Therefore,
\begin{align}
&\sum_{i=1}^F \sum_{q=1}^Q M_q p_{i,q} = F \cdot \sum_{q=1}^Q M_q, \nonumber \\
&~~~\text{if } \nu \leq \min \Big\{ \frac{f_i \mathcal{P}(q) C_{i,q} }{M_q}e^{-C_{i,q}},~\forall i,~\forall q \Big\}.
\end{align}
However, if $\nu \geq \max\{ \frac{f_i \mathcal{P}(q) C_{i,q}}{M_q},~\forall i \in \mathcal{F},~\forall q \in \mathcal{Q} \}$, $p_{i,q} = 0,~\forall i \in \mathcal{F},~\forall q \in \mathcal{Q}$, and 
\begin{equation}
\sum_{i=1}^F \sum_{q=1}^Q M_q p_{i,q} = 0,~\text{if } \nu \leq \max \Big\{ \frac{f_i \mathcal{P}(q) C_{i,q}}{M_q},~\forall i,~\forall q \Big\}.
\end{equation}
Thus, if $\min\{ \frac{f_i \mathcal{P}(q) C_{i,q}}{M_q} e^{- C_{i,q}},~\forall i,~\forall q \} \leq \nu \leq \max\{ \frac{f_i \mathcal{P}(q) C_{i,q}}{M_q},~\forall i,~\forall q \}$, 
\begin{equation}
0 \leq \sum_{i=1}^F \sum_{q=1}^Q M_q p_{i,q} \leq F \cdot \sum_{q=1}^Q M_q.
\end{equation}

Assuming that $M < F \cdot \sum_{q=1}^Q M_q$, since $\sum_{i=1}^F \sum_{q=1}^Q M_q p_{i,q}$ is decreasing with $\nu$, we can find the optimal $\nu^*$ and $p_{i,q}^*$ by the bi-section method.
The details of the bi-section method for optimal file placement are shown in Algorithm \ref{algo:node_association}.

\begin{algorithm}
	\caption{Bisection method for optimal file placement rule
		\label{algo:file_placement}}
	\begin{algorithmic}[1]
		\State{Initialize $\epsilon, \nu_{-} = \min\{ l_{i,q}, \forall i \in \mathcal{F}, \forall q \in \mathcal{Q} \}$, $\nu_{+} = \max\{ u_{i,q}, \forall i \in \mathcal{F}, \forall q \in \mathcal{Q} \}$, \\~~~~~~~~~~ and 				$p_{i,q}^* = -1, \forall i \in \mathcal{F}, \forall q \in \mathcal{Q}$}
		\Comment{$\epsilon$: error tolerance threshold}
		\While{$| \sum_{i=1}^F \sum_{q=1}^Q M_q p_{i,q}^* - M | \geq \epsilon$}{
			\State{$\nu^* = (\nu_{-}+\nu_{+})/2$}
			\State{$\mu_{i,q}^*=[ f_i \mathcal{P}(q) C_{i,q} e^{-C_{i,q}} - \nu^* M_q ]^{+},~\forall i \in \mathcal{F},~\forall q \in \mathcal{Q}$}
			\State{$p_{i,q}^* = \frac{1}{C_{i,q}} \Big[ \log_2(f_i \mathcal{P}(q) C_{i,q}) - \log_2(\nu^* M_q + \mu_{i,q}^*) \Big]^{+},~\forall i \in \mathcal{F},~\forall q \in \mathcal{Q}$}
			\If{$\sum_{i=1}^F \sum_{q=1}^Q M_q p_{i,q}^* > M$}
			$\nu_{-} \leftarrow \nu^*$
			\ElsIf{$\sum_{i=1}^F \sum_{q=1}^Q M_q p_{i,q}^* < M$}
			$\nu_{+} \leftarrow \nu^*$
			\EndIf
		}
		\EndWhile
	\end{algorithmic}
\end{algorithm}

\section{Node Association Maximizing Video Quality with Playback Delay Constraint}
\label{sec:node_association}

The node association problem in this paper amounts to choosing the cache-enabled devices for $N$ users to request video files.
After making the candidate set of devices which caches the requested file, the user has to choose the specific device as well as the level of quality.
This paper proposes a dynamic algorithm for users to associate with cache-enabled devices to maximize time-average video quality measures with a playback delay constraint.
Improvement in video playback latency can be explained based on the user queue model.

\subsection{User Queue Model}

A video file consists of many sequential chunks.
User terminals receive video files from cache-enabled devices and process data for video streaming services in units of chunks.
Each chunk of a file is responsible for some playback time of the entire stream.
As long as all chunks are in correct sequence, each chunk can have different quality in dynamic streaming. 
Therefore, users can dynamically choose video quality levels in every chunk processing time.
By using the queue model, it can be said that the playback delay occurs when the chunk to be played does not yet arrive at the queue. 
In this sense, receiver queue dynamics collectively reflects the various factors which cause the playback delay.

In general, user queue models have their own arrival and departure processes. 
For each user $n \in \{1,\cdots,N\}$, the queue dynamics in each time slot $t\in \{0,1,\cdots,\}$ can be represented as follows:
\begin{eqnarray}
Q_n[t+1] &=& \max \{ Q_n[t]-b_n[t], 0 \} + a_n[t] \label{eq:Q_dynamics1}\\
Q_n[0] &=& 0 \label{eq:Q_dynamics2}
\end{eqnarray}
where $Q_n[t]$, $a_n[t]$, and $b_n[t]$ stand for the queue backlog, the arrival and departure processes of user $n$ at time $t$, respectively.  
The queue states are updated and every user performs node association in each unit time slot $t$. 
In this paper, the interval of each slot is determined to be the channel coherence time, $\tau_c$. 
Suppose a block fading channel, whose channel gain is static during the processing of multiple chunks, $t_c = m \tau$, where $\tau$ is a chunk processing time and $m$ is the positive integer.

In this paper, queue backlog $Q_n[t]$ counts the number of video chunks in the queue. 
$a_n[t]$ and $b_n[t]$ semantically mean the numbers of received and processed chunks.
Simply, $m$ chunks are processed in each time slot, so $b_n[t]=m$.
On the other hand, $a_n[t]$ obviously depends on the data rate of the communication link between user $n$ and its associated device and the chunk size.
The departure and arrival processes are given as follows:
\begin{eqnarray}
a_n[t] &=& \bigg\lfloor \frac{R_n(\alpha_n(t),t)\cdot \tau_c}{L(q_n(\alpha_n(t),i_n),t)} \bigg \rfloor \label{eq:arrival}\\
b_n[t] &=& m
\end{eqnarray}
where $\alpha_n(t)$ denotes the cache-enabled device associated with user $n$ at time $t$, and $q_n(\alpha_n(t),i_n)$ is the quality level of file $i_n$ which user $n$ requests from the device $\alpha_n(t)$.
Also, $R_n(\alpha_n(t),t)$ and $L(q_n(\alpha_n(t),i_n),t)$ indicate the data rate of a D2D link between user $n$ and the device $\alpha_n(t)$, and a chunk size of file $i_n$ of the desired quality $q_n(\alpha_n(t),i_n)$ at time $t$, respectively.
Some video chunks can be only partially delivered as the channel condition varies and node association is updated at every time slot $t$.
Since partial chunk transmission is meaningless in our algorithm, the flooring is used in (\ref{eq:arrival}). 

Let the Rayleigh fading channel between user $n$ and device $\alpha_n(t)$ denoted by $h_{n,t}$.
Then, the link rate between user $n$ and device $\alpha_n(t)$ is simply given by
\begin{equation}
R_n(\alpha_n(t),t) = \mathcal{B}\log_2 \Big( 1+ \frac{|h_{n,t}|^2}{\sigma^2} \Big).
\label{eq:data_rate}
\end{equation}

For video streaming service, it is important to avoid playback delay.
The user needs a chunk in the next sequence during video playback. 
If the next chunk has not yet arrived in the queue, there will be a delay in playback.
Therefore, stacking enough queue backlogs, i.e., video chunks in sequence, is necessary for averting playback delay.
Suppose that the queue is almost empty.
In this case, the cache-enabled device whose channel is strong and which stores the requested file of low quality (i.e., small chunk size) is preferable for the user.
On the other hand, when the queue is filled with a lot of video chunks, the user can request the high-quality video file without worrying about playback delay. 

\textit{Remark}: If $\tau_c$ is too long, it is better to update node associations more frequently than channel variations.
For example, consider a user whose queue is filled with many chunks and thus is associated with the caching device delivering the chunks smaller than $b_n[t]=m$.
If this situation persists for a long time, chunks in the queue will be emptied out soon and playback delay will occur, therefore several updates of node association are required over the time interval of $\tau_c$.
On the other hand, if $\tau_c$ is too short, the requested video cannot be successfully delivered even when the data rate of the link is good, because of the flooring in (\ref{eq:arrival}).
Therefore, block fading is assumed with $\tau_c$ large enough for the users to receive the video chunks.

\begin{figure} [h!]
	\centering
	\includegraphics[width=0.4\textwidth]{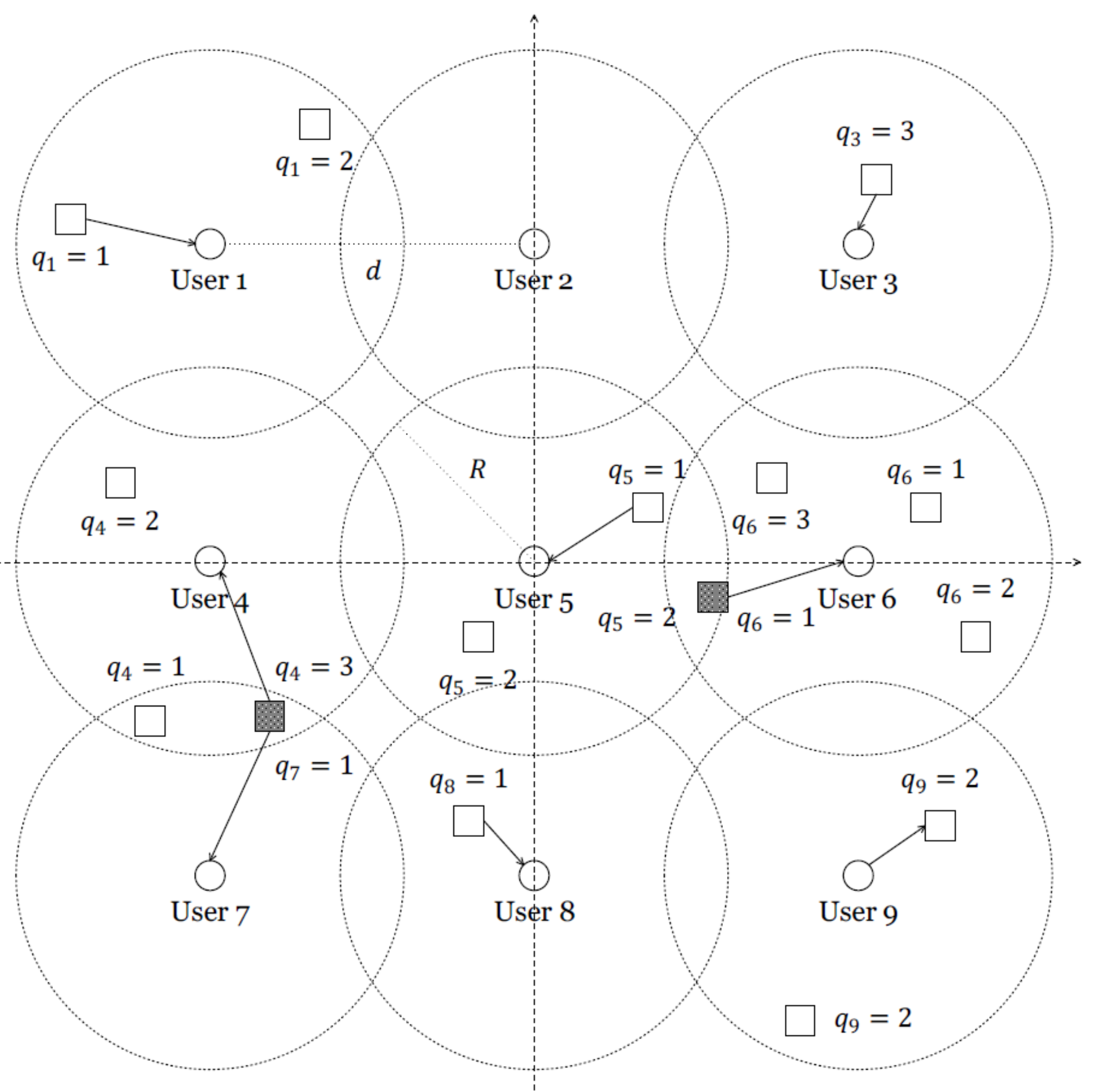}
	\caption{Node Association Cases}
	\label{Fig:NodeAssociationCases}
\end{figure}

\subsection{Node Association Cases}

Depending on geological locations of cache-enabled devices, node association of certain user with an appropriate cache-enabled device can be classified into a number of cases.
These example cases are illustrated in Fig. \ref{Fig:NodeAssociationCases}.
Fig. \ref{Fig:NodeAssociationCases} assumes that each user requests the video file from one of cache-enabled devices within radius $R$, and there are quality levels of 1, 2, and 3.
Only the devices which cache the requested file are depicted in Fig. \ref{Fig:NodeAssociationCases}. 
The quality levels of requested videos in cache-enabled devices are written as $q_n = c,~c\in\{1,2,3\}$, to indicate that the device caches the video of quality level $c$ requested by user $n$. 
In particular, the devices which receive multiple file delivery requests are shown as the shaded squares.
The proposed dynamic algorithm for node association can be applied to cases 4, 5, 6, and 7.

\begin{itemize}
	\item Case 1:
	When there is no caching device which caches the requested file within radius $R$ of the user, the user should download the video file from the BS via a cellular link. (user 2)
	
	\item Case 2:
	When there is only one device caching the requested file within radius $R$ of the user, the user just downloads the video from this device, but only the fixed quality can be provided. 
	If the user wants the high-quality file, it can download file from the BS but this option is not considered in this paper. (user 3, 8)
	
	\item Case 3:
	When there are multiple devices caching the requested files of the same quality within radius $R$ of the user, the user requests the video from one of the devices whose channel is the strongest. (user 9)
	Similar to Case 2, only the fixed quality can be provided. 
	
	\item Case 4:
	When there are multiple devices caching the requested files of different quality levels within radius $R$ of the user, the proposed dynamic algorithm can be applied for node association. 
	The proposed algorithm maximizes the expected video quality constrained on sufficiently large queue backlog to avoid playback delay. (user 1)
	
	\item Case 5:
	When the cache-enabled device receives two or more file delivery requests including the target user's, i.e., \textit{request collision} occurs, the device serves multiple users by NOMA. (user 4, 7)
	The proposed algorithm determines to whether to exploit NOMA.
	
	\item Case 6:
	When the cache-enabled device receives two or more file delivery requests including the target user's, 
	the proposed algorithm makes the device to schedule the target user and to ignore other requests. (user 6)  
	
	\item Case 7:
	When the cache-enabled device receives two or more file delivery requests including the target user's,
	the proposed algorithm determines the device to schedule another user and to ignore the request of the target user. 
	Then, the target user should find another cache-enabled device, and if there is no other device which stores the requested file within radius $R$ , it has to download the file from the BS. (user 5)
\end{itemize}

\subsection{Dynamic Node Association for Video File Delivery under Queue Stability}
\label{subsec:node_association}

We specifically go after the following optimization problem:
\begin{align}
\text{max.}&~\displaystyle\sum_{n \in \mathcal{N}} \mathbb{E} [ \mathcal{P}(q_n (\alpha_n(t),i_n) ) ]
\label{eq:optProb1}\\
\text{s.t.}&~\underset{t \rightarrow \infty}{\lim} \frac{1}{t} \sum_{t'=0}^{t-1} \mathbb{E}[Z_n[t']] < \infty,~\forall n\in \mathcal{N}
\label{eq:optConst}
\end{align}
where $\mathcal{N}$ is the set of $N$ file-requesting users via D2D links, 
and $Z_n[t] = \tilde{Q} - Q_n[t]$. 
The optimization metric (\ref{eq:optProb1}) is the sum of the time averaged video quality measures of the file-requesting users as given by
\begin{equation}
\sum_{n \in \mathcal{N}} \Big[ \underset{t \rightarrow \infty}{\lim} \frac{1}{t} \sum_{t'=0}^{t-1} \mathcal{P}(q_n(\alpha_n(t'),i_n)) \Big].
\end{equation}

Here, $Z_n[t]$ is introduced to make $Q_n[t]$ large enough to avoid playback delay, and $\tilde{Q}$ is a sufficiently large parameter which affects the maximal queue backlog.
From (\ref{eq:Q_dynamics1}) and (\ref{eq:Q_dynamics2}), the queue dynamics of $Z_n[t]$ can be represented as follows:
\begin{eqnarray}
Z_n[t+1] &=& \min \{ Z_n[t]+b_n[t], \tilde{Q} \} - a_n[t] \label{eq:Z_dynamics1}\\
Z_n[0] &=& \tilde{Q} \label{eq:Z_dynamics2}.
\end{eqnarray}

Even though the update rules of $Q_n[t]$ and $Z_n[t]$ are different, both queue dynamics mean the same video chunk processing.
Therefore, playback delay due to emptiness of $Q_n[t]$ can be explained by queuing delay of $Z_n[t]$.
By Little's theorem \cite{Book:DataNetworks}, the expected value of $Z_n[t]$ is proportional to the time-averaged queuing delay. 
Therefore, we hope to limit the queuing delay by addressing the constraint (\ref{eq:optConst}), and it is well known that Lyapunov optimization with the constraint (\ref{eq:optConst}) can make $Z_n[t]$ bounded \cite{Book:Neely}.

Let $\mathbf{Z}[t]$ denote the column vector of $Z_n[t]$ of all users at time $t$, and define the quadratic Lyapunov function $L(\mathbf{Z}[t])$ as follows:
\begin{equation}
L(\mathbf{Z}[t]) = \frac{1}{N}\sum_{n\in \mathcal{N}} (Z_n [t])^2
\end{equation}
Then, let $\Delta (.)$ be a conditional quadratic Lyapunov function that can be formulated as $\mathbb{E}[L(\mathbf{Z}[t+1])-L(\mathbf{Z}[t]) | \mathbf{Z}[t]]$, i.e., the drift on $t$.
The dynamic policy is designed to solve the given optimization problem (\ref{eq:optProb1}) by observing the current queue state, $Z_n[t]$, and determining the node association to minimize a upper bound on \textit{drift-plus-penalty} \cite{TC2015bethanabhotla}:
\begin{equation}
\Delta (\mathbf{Z}[t]) - \tilde{V} \mathbb{E}\bigg[\sum_{n\in \mathcal{N}} \mathcal{P}(q_n(\alpha_n(t),t)) | \mathbf{Z}[t]\bigg].
\label{eq:drift-plus-penalty}
\end{equation}

At first, find the upper bound on the change in the Lyapunov function.
\begin{align}
&L ( \mathbf{Z}[t+1] ) - L ( \mathbf{Z}[t] ) = \frac{1}{N} \sum_{n\in \mathcal{N}} \Big[ Z_n[t+1]^2 - Z_n[t]^2 \Big]  \\
&= \frac{1}{N} \sum_{n\in \mathcal{N}} \Big[ Q_n[t+1]^2 - Q_n[t]^2 - 2\tilde{Q} (Q_n[t+1] - Q_n[t]) \Big]  \\
&= \frac{1}{N} \sum_{n\in \mathcal{N}} \Big[ (\max[Q_n[t] - b[t],0] + a_n[t])^2 - Q_n[t]^2 \nonumber \\
&~~~~~~~- 2\tilde{Q} ( \max[Q_n[t] - b_n[t],0] + a_n[t] - Q_n[t] ) \Big]  \\
&\leq \frac{1}{N} \sum_{n\in \mathcal{N}} \Big[ b_n[t]^2 + a_n[t]^2 - 2Q_n[t] b_n[t] \nonumber \\ 
&~~~~~~~- 2(\tilde{Q}-Q_n[t]) a_n[t] + 2\tilde{Q} Q_n[t] \Big] 
\end{align}
Then, the upper bound on the conditional Lyapunov drift is obtained as
\begin{align}
&\Delta (\mathbf{Z}[t]) = \mathbb{E} [ L ( \mathbf{Z}[t+1] ) - L ( \mathbf{Z}[t] ) | \mathbf{Z}[t]] \\
&\leq \frac{1}{N} \sum_{n\in \mathcal{N}} \Big[1 - 2Q_n[t] + 2\tilde{Q}Q_n[t] \Big] \nonumber \\
&~~ + \mathbb{E} \bigg[ \frac{1}{N} \sum_{n\in \mathcal{N}} a_n[t]^2 \Big| \mathbf{Z}[t] \bigg] \nonumber \\
&~~ - \mathbb{E} \bigg[ \frac{1}{N} \sum_{n\in \mathcal{N}}  2 (\tilde{Q} - Q_n[t]) \cdot a_n[t] \Big| \mathbf{Z}[t] \bigg].
\end{align}
According to (\ref{eq:drift-plus-penalty}), minimizing a bound on \textit{drift-plus-penalty} is consistent with minimizing 
\begin{align}
&\mathbb{E} \bigg[ \frac{1}{N} \sum_{n\in \mathcal{N}}  a_n[t]^2 \Big| \mathbf{Z}[t] \bigg] - \tilde{V} \mathbb{E} \bigg[ \sum_{n\in \mathcal{N}} \mathcal{P}(q_n(\alpha_n(t),t)) \Big| \mathbf{Z}[t] \bigg] \nonumber \\
&~- \mathbb{E} \bigg[ \frac{1}{N} \sum_{n\in \mathcal{N}}  2 (\tilde{Q} - Q_n[t]) \cdot a_n[t] \Big| \mathbf{Z}[t] \bigg].
\label{eq:drift-plus-penalty-expectation}
\end{align}
We now use the concept of opportunistically minimizing the expectations, so (\ref{eq:drift-plus-penalty-expectation}) is minimized by the algorithm which observes the current queue state, $\mathbf{Z}[t]$ (i.e., $\mathbf{Q}[t]$ given $\tilde{Q}$) and chooses $\alpha_n(t)$ for all $n\in \mathcal{N}$ to minimize
\begin{align}
&\sum_{n\in \mathcal{N}}  a_n[\alpha_n(t),t]^2 - V \sum_{n\in \mathcal{N}} \mathcal{P}(q_n(\alpha_n(t),t)) \nonumber \\
&~- \sum_{n\in \mathcal{N}}  2 (\tilde{Q}-Q_n[t]) \cdot a_n[\alpha_n(t),t],
\label{eq:minimizing_metric}
\end{align}
where $V = \tilde{V}\cdot N$ and $a_n[t]$ is replaced by $a_n[\alpha_n(t),t]$ to emphasize the decision parameter of $\alpha_n(t)$.

From (\ref{eq:minimizing_metric}), we can anticipate how the algorithm works. 
When the queue of user $n$ is almost empty, the large arrivals are necessary for user $n$ not to wait the next video chunk.
In this case, user $n$ prefers the device which gives many arrivals.
On the other hand, when the queue backlogs are stacked enough to avoid playback delay, $Q_n(t) \simeq \tilde{Q}$, user $n$ requests the video of high quality without worrying about playback latency.

System parameter $V$ in (\ref{eq:minimizing_metric}) is a weight factor for the term representing video quality measure. The relative value of $V$ to $\tilde{Q}-Q_n(t)$ is important to control the queue backlogs and quality measures at every time.
The appropriate initial value of $V$ needs to be obtained by experiment because it depends on the distribution of the cache-enabled devices, the channel environments, and the threshold of queue backlog, $\tilde{Q}$.
Also, $V\geq 0$ should be satisfied. 
If $V<0$, users prefer low-quality videos even when a lot of video chunks have already arrived at the user queue.
Moreover, in the case of $V=0$, the user only aims at stacking queue backlogs without consideration of video quality.
On the other hand, when $V\rightarrow \infty$, users do not consider the queue state, and thus they just request the highest-quality files.
$V$ can be regarded as the parameter to control the trade-off between video quality and playback delay.

Since streaming users cannot know other users' channel gains, each user independently finds the cache-enabled device which stores the video file of desired quality.
Therefore, (\ref{eq:minimizing_metric}) is treated separately, and each user minimizes its own objective function:
\begin{align}
g_n(\alpha_n(t),t) &= a_n[\alpha_n(t),t]^2 - V \mathcal{P}(q_n(\alpha_n(t),t)) \nonumber \\
&~-  2 (\tilde{Q}-Q_n[t]) \cdot a_n[\alpha_n(t),t].
\label{eq:minimizing_metric_indep}
\end{align}
Since there is a finite number of cache-enabled devices within the radius $R$ of the user, each user can easily find the device for video delivery, i.e., determination of $\alpha_n(t)$, by greedy search.

However, if two or more users simultaneously request files from the same cache-enabled device, the objective functions of those users are not independent. 
The reason is that the data rates of the users are obtained for one-to-one communication, (\ref{eq:data_rate}), but the device which receives multiple file requests cannot provide the data rate of (\ref{eq:data_rate}) to all file-requesting users.
We shall call this situation the \textit{request collision}.
Since the cache-enabled device which experiences request collision can receive channel information of all file-requesting users from them, the device should resolve request collision by jointly minimizing the sum of objective functions of those users.

Assume that there are $J$ user sets, $\mathcal{N}_{rc}(j),~j=1,\cdots,J$, whose element users request files from the same device.
Note that $\alpha_n(t)$ is the same for all $n \in \mathcal{N}_{rc}(j)$. 
Let $\mathcal{N}_{rc} = \mathcal{N}_{rc}(1) \cup \mathcal{N}_{rc}(2) \cup \cdots \cup \mathcal{N}_{rc}(J)$.
Then, (\ref{eq:minimizing_metric}) can be re-written as
\begin{align}
&\underset{n\in \mathcal{N}- \mathcal{N}_{rc}}{\sum} g_n(\alpha_n(t),t) + \sum_{j=1}^{J} \underset{n\in \mathcal{N}_{rc}(j)}{\sum} g_n(\alpha_n(t),t). 
\label{eq:metric_rc1}
\end{align}

The first term of (\ref{eq:metric_rc1}) is separable, so each user $n \in \mathcal{N}-\mathcal{N}_{rc}$ just minimizes its own objective function of (\ref{eq:minimizing_metric_indep}).
Likewise, the summations over users $n \in \mathcal{N}_{rc}(j)$ for different $j$ are also separable, so we can independently minimize 
\begin{equation}
\underset{n\in \mathcal{N}_{rc}(j)}{\sum} g_n(\alpha_n(t),t) \label{eq:minimize_metric_rc}
\end{equation}
for every $j=1,\cdots,J$.
However, the element terms of summation over certain user set $\mathcal{N}_{rc}(j)$ are not independent, so additional steps are necessary to handle the occurrence of request collisions.
There are two solutions: 1) scheduling of one user minimizing objective function (\ref{eq:minimize_metric_rc}) and 2) NOMA to response to the multiple requests simultaneously. 

\subsection{Approaches Against Request Collision}

\subsubsection{Scheduling of One User Minimizing Objective Function}
In this approach, the cache-enabled device at which request collision occurs simply schedules one of the file-requesting users for video delivery, by minimizing the value of (\ref{eq:minimize_metric_rc}).
After scheduling of only one user, say user $n_0$, others find another cache-enabled devices, $\alpha'_n(t), \forall n\in \mathcal{N}_{rc}(j), n\neq n_0$, within radius $R$ of each user, separately. 
For the choices of $\alpha'_n(t)$, users follow the steps of Section \ref{subsec:node_association}, without the consideration of the cache-enabled device chosen at first, $\alpha_n(t)$.
If there is no device for video delivery except for $\alpha_n(t)$, then this user should request the file from the BS.

Then, the caching device at which request collision occurs, $\alpha_n(t)$, computes $|\mathcal{N}_{rc}(j)|$ metrics of (\ref{eq:minimize_metric_rc}) for every case of scheduling of user $n \in \mathcal{N}_{rc}(j)$ and finds the one giving the minimum value. 
Thus, a choice of user $n_0$ can be obtained by 
\begin{equation}
n_0 = \underset{n \in \mathcal{N}_{\text{rc}}(j)}{\arg \min}~ g_n(\alpha_n(t),t) + \sum_{\substack{m \in \mathcal{N}_{rc}(j)\\ m\neq n}} g_m(\alpha'_m(t),t) \label{eq:scheudlingUser},
\end{equation}
and let the minimum value denoted by $\mathcal{M}_O(j)$:
\begin{equation}
\mathcal{M}_O(j) = g_{n_0}(\alpha_{n_0}(t),t) + \underset{\substack{m\in \mathcal{N}_{rc}(j)\\ m \neq n_0}}{\sum} g_m(\alpha'_m(t),t). 
\end{equation}

Unfortunately, scheduling of one user could have a serious problem that conflicts with the noise-limited constraint, which does not allow the additional D2D link within the radius $R$ of the streaming user whose D2D link is already constructed.
If there are large overlaps among users' coverages of radius $R$, the cache-enabled device which is newly found by the unscheduled user $n$, $n \neq n_0$, would be in the coverage of the user $n_0$. 
If so, this newly found link cannot be activated, and the unscheduled user should find another device again or directly receive the file from the BS.
Furthermore, when $\lambda$ is small, i.e., cache-enabled devices are sparsely located, it is likely that the unscheduled users cannot find another neighboring device. 
To combat these problems, NOMA is proposed to handle the multiple requests simultaneously. 
Since receiving the file from the neighboring device is much more advantageous in terms of transmission latency than downloading from the BS via a cellular link, NOMA would be preferred in above cases.

\subsubsection{NOMA}
The cache-enabled device can respond to multiple file requests simultaneously by employing NOMA. 
Although the NOMA signals transmitted to users interfere with each other, an advanced receiver, e.g., successive interference cancellation (SIC), can successfully remove interference \cite{tse}.
However, since multiple users are served within the same resource in NOMA, degradations of data rates are inevitable.
Therefore, NOMA would be useful if the system prefers to guarantee reduced transmission latency at the expense of data rate degradation. 

When the cache-enabled device utilizes power-multiplexing NOMA, different power ratios, $\beta=[\beta_{m_{j,1}},\cdots,\beta_{m_{j,|\mathcal{N}_{rc}(j)|}}]$, are weighted on the signals of all users, $m_{j,l} \in \mathcal{N}_{rc}(j),~l\in \{ 1,\cdots,|\mathcal{N}_{rc}(j)| \}$. 
Larger power is usually allocated to the user which experiences the weaker channel condition, so power allocation ratios for file-requesting users satisfy $\beta_{m_{j,1}} < \beta_{m_{j,2}} < \cdots < \beta_{m_{j,|\mathcal{N}_{rc}(j)|}}$, with the assumption that $|h_{m_{j,1},t}|^2 > |h_{m_{j,2},t}|^2 > \cdots > |h_{m_{j,|\mathcal{N}_{rc}|},t}|^2$.
The data rate of user $m_{j,l} \in \mathcal{N}_{rc}(j),~l\in \{ 1,\cdots,|\mathcal{N}_{rc}(j)| \}$ in NOMA system is given by \cite{tse}
\begin{equation}
R_n^N(\alpha_n(t),t) = \mathcal{B}\log_2\bigg( 1+ \frac{|h_{m_{j,l},t}|^2 \beta_{m_{j,l}}}{ |h_{m_{j,l},t}|^2 \sum_{l'=1}^{l-1} \beta_{m_{j,l'}} + \sigma^2 } \bigg).
\label{eq:NOMArate}
\end{equation}
The data rate of (\ref{eq:NOMArate}) can be obtained by performing SIC for the signals of the users with weaker channels than user $m_{j,l}$.
In this case, as $N$ increases, data rates of all file-requesting users are significantly degraded. 
The objective function for users $n \in \mathcal{N}_{rc}(j)$ is changed as follows:
\begin{equation}
\mathcal{M}_N(j) = \underset{n\in \mathcal{N}_{rc}(j)}{\sum} g_n^N(\alpha_n(t),t),
\end{equation}
where $g^N_n(\alpha_n(t),t)$ is obtained by substituting $R^N_n(\alpha_n(t),t)$ for $R_n(\alpha_n(t),t)$ in (\ref{eq:minimizing_metric_indep}).

Finally, we decide which approach is better to handle the request collision for each user set, $\mathcal{N}_{rc}(j)$ for all $j=1,\cdots,J$, by comparing $\mathcal{M}_N(j)$ with $\mathcal{M}_O(j)$.
If $\mathcal{M}_N(j) > \mathcal{M}_O(j)$, scheduling of one user is better than NOMA but, otherwise, NOMA is preferred.

\begin{algorithm}[t!]
	\caption{Dynamic node association for maximization of time-average video streaming quality sum
		\label{algo:node_association}}
	\begin{algorithmic}[1]
		\Require{\\
			\begin{itemize}
				\item $V$: parameter for streaming quality-delay trade-offs
				\item $\tilde{Q}$: threshold for queue backlog size
			\end{itemize}
		}
		\State{$t=0$ // $T$: number of discrete-time operations}
		\While{$t \leq T$}{
			\State{Observe $Q_{n}[t]$}
			\State{For users $n\in \mathcal{N}$, associate with the cache-enabled device, 
				$\alpha_n^*(t) = \underset{\alpha_n(t)}{\arg \min}$ (\ref{eq:minimizing_metric_indep}).}
			\State{Find $\mathcal{N}_{rc}(j),~j=1,\cdots,J$.}
				\For{$j=1:J$}{
					\State{Compute $\mathcal{M}_N(j)$ and $\mathcal{M}_O(j)$, 
						\newline 
						$~~~~~~~~~~$and find $n_0$ and $\alpha'_n(t),~\forall n\in \mathcal{N}_{rc}(j),~n\neq n_0$.}
					\If{$\mathcal{M}_N (j) > \mathcal{M}_O (j)$}
					\State{$\alpha^*_{n_0}(t) = \alpha_n(t)$}
					\State{$\alpha^*_n(t) = \alpha'_n(t),~\forall n\in \mathcal{N}_{rc}(j),~n\neq n_0$}
					\EndIf
				}
				\EndFor
		}
		\EndWhile
	\end{algorithmic}
\end{algorithm}

\section{Performance Evaluation}
\label{sec:performance_evaluation}

In this section, we show that the proposed algorithms for file placement and node association work well with video files of different quality levels.
We set the parameters, $F=5$, $Q=3$, and $M=6$.
Also, we assume that $\gamma=1$ and $\rho_{i,q} = \mathcal{B},~\forall i,~\forall q$.
PSNR is considered as a video quality measure, and according to \cite{ICTC2015kim}, quality measures and file sizes depending on quality levels are $\mathcal{P}(q)=[34,~ 36.64,~ 39.11]$ dB and $L(q) = [2621,~5073,~10658]$ kbits, respectively.
Especially for finding the optimal caching probabilities, the approximately normalized file size $M_q = [1,~2,~4]$ is used.

\begin{table*}[t!]%
	\caption{Optimal Caching Probabilities with $\lambda=0.1$ and SNR=20dB, when $M_q(3)=4$ ($M_q(3)=3,~M_q(3)=6$)}
	\label{Table:CachingProb}
	{\footnotesize
		\begin{center}
			\begin{supertabular}{c|ccc}
				\toprule
				& \multicolumn{3}{c}{Quality level} \\
				\cmidrule{2-4}
				File type    & 1 & 2 & 3 \\
				\midrule
				1 & 0.2222 (0.2438, 0.1972) & 0.2183 (0.2399, 0.1932) & 0.2126 (0.2474, 0.1689) \\
				2 & 0.1904 (0.2120, 0.1653) & 0.1865 (0.2080, 0.1614) & 0.1807 (0.2155, 0.1371) \\
				3 & 0.1717 (0.1933, 0.1467) & 0.1678 (0.1894, 0.1428) & 0.1621 (0.1969, 0.1185) \\
				4 & 0.1585 (0.1801, 0.1335) & 0.1546 (0.1762, 0.1296) & 0.1489 (0.1837, 0.1052) \\
				5 & 0.1483 (0.1699, 0.1233) & 0.1444 (0.1660, 0.1193) & 0.1387 (0.1735, 0.0950) \\
				\bottomrule
			\end{supertabular}
		\end{center}
	}
\end{table*}

\subsection{Optimal Caching Probabilities and Effects of Storage Size, Device Intensity, and SNR} 

According to (\ref{eq:optCachingProb}), the optimal caching probabilities depend on $\lambda$ and SNR.
As an example, the optimal caching probabilities with $\lambda=0.1$ and SNR = 20dB are shown in Table \ref{Table:CachingProb}. 
In Table \ref{Table:CachingProb}, the caching probability of the popular and low-quality file is larger than that of the unpopular and high-quality file.
However, caching probabilities for different quality levels are not much different in this system, and this means that the trade-off between video quality and video diversity is unbiased.
Actually, this trade-off depends on the relative values of the quality measures to the file sizes. 
If we arbitrarily change the file size of the quality level 3 with the fixed quality measure value, different caching probabilities are obtained.
In Table \ref{Table:CachingProb}, all the first values in parentheses are for $M_q(3)=3$ and the second values are for $M_q(3)=6$, rather than $M_q(3)=4$.
When the file size of quality level 3 reduces to $M_q(3)=3$, the relative file size to the quality measure decreases also, so all the caching probabilities of files of quality level 3 increase.
On the other hand, when the file size of quality level 3 increases to $M_q(3)=6$, the differences of caching probabilities between quality 1 and quality 3 increase, compared to when $M_q(3)=4$.

\begin{figure} [h!]
	\centering
	\includegraphics[width=0.37\textwidth]{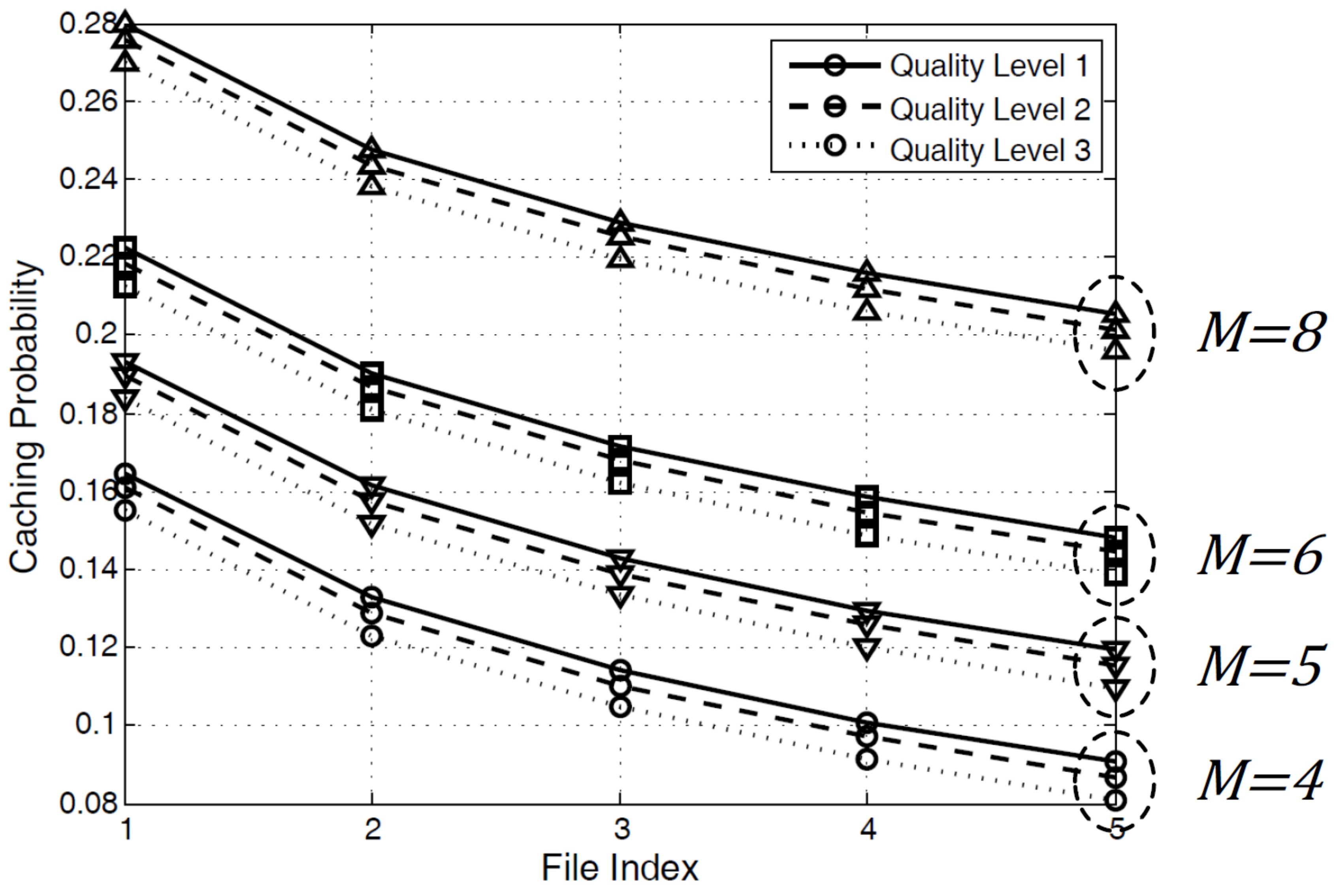}
	\caption{Caching probabilities with different values of $M$}
	\label{Fig:CachingProb_vs_FileIndex}
\end{figure}

Fig. \ref{Fig:CachingProb_vs_FileIndex} gives the plots of caching probabilities versus file indices with different storage sizes, $M$, assuming $\lambda=0.1$ and SNR = 20dB. 
As $M$ grows, all caching probabilities increase almost linearly. 
However, the differences among the caching probabilities with different quality levels are not changed, because they are influenced by the relative values of the quality measures to file sizes, as explained above.

\begin{figure} [h!]
	\centering
	\includegraphics[width=0.37\textwidth]{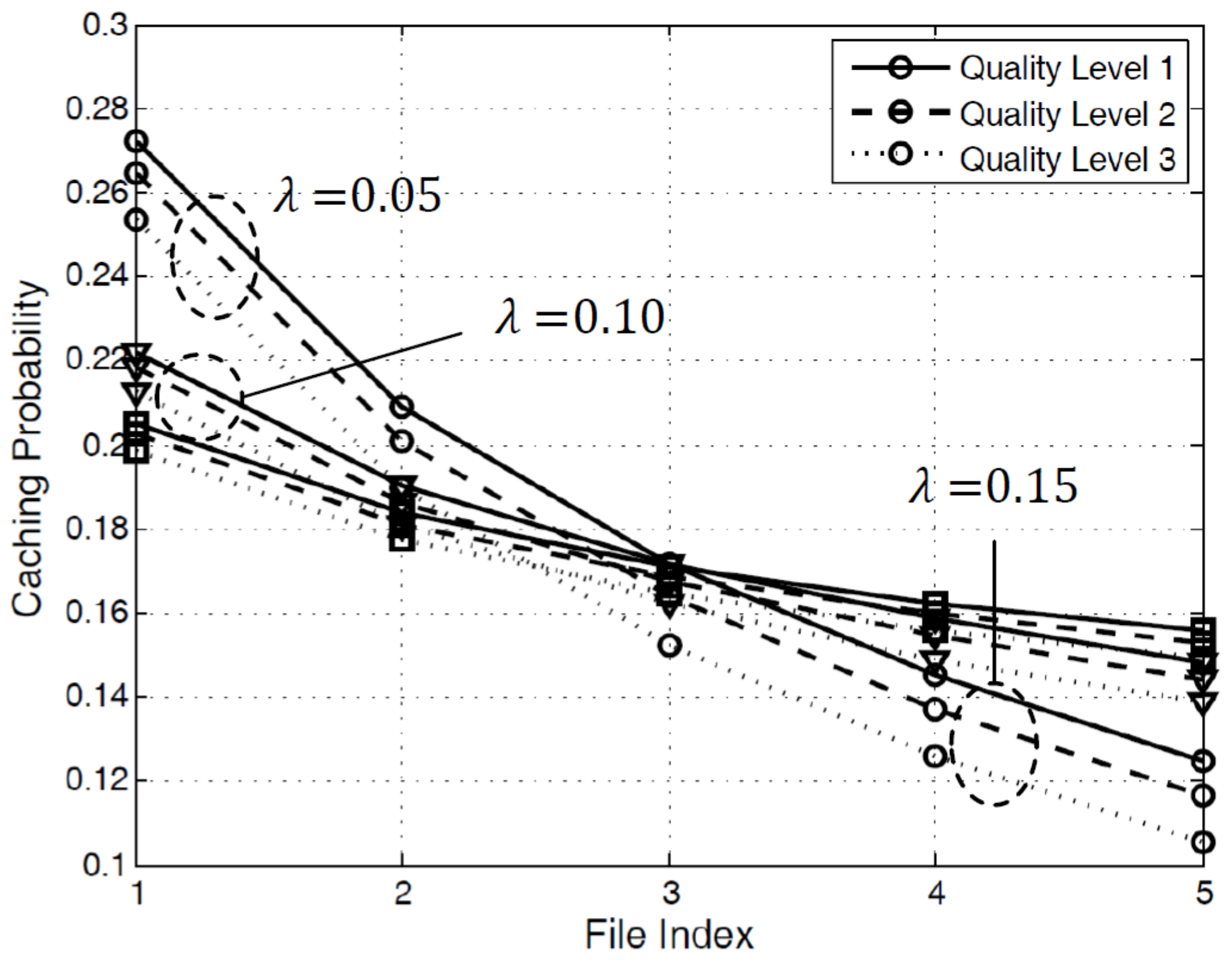}
	\caption{Caching probabilities with different values of $\lambda$}
	\label{fig:CachingProb_vs_FileIndex_lambda}
\end{figure}

\begin{figure} [h!]
	\centering
	\includegraphics[width=0.37\textwidth]{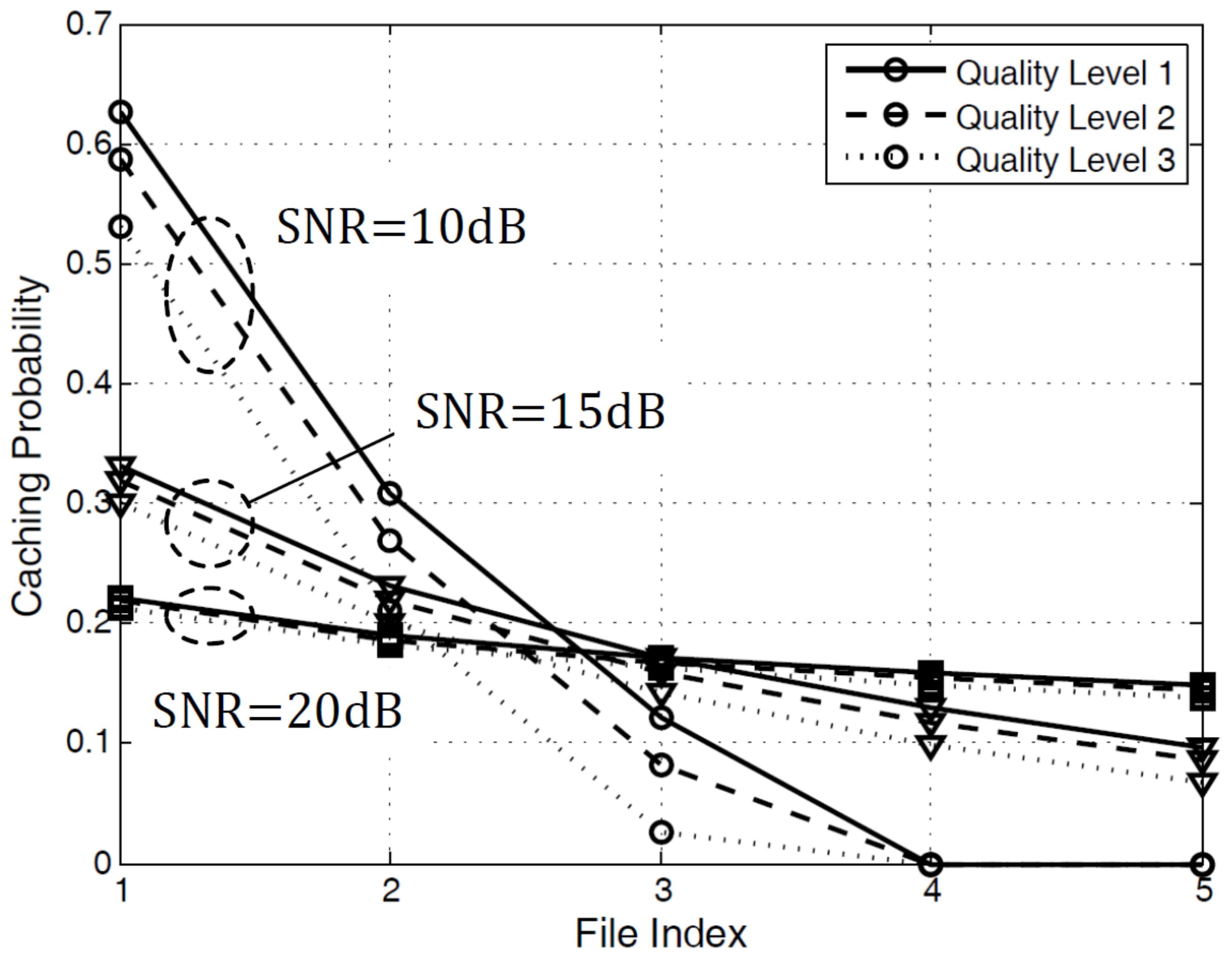}
	\caption{Caching probabilities with different SNRs}
	\label{fig:CachingProb_vs_FileIndex_SNR}
\end{figure}

Fig. \ref{fig:CachingProb_vs_FileIndex_lambda} and \ref{fig:CachingProb_vs_FileIndex_SNR} show the effects of $\lambda$ and SNR on caching probabilities.
Here, smaller $\lambda$ and smaller SNR give the similar effects, i.e., both make file delivery via a D2D link difficult. 
Smaller $\lambda$ means that there are a smaller number of cache-enabled devices within radius $R$ of the user, and a smaller SNR makes the successful file delivery more difficult.
Therefore, when $\lambda$ and/or SNR are small, caching probabilities become biased to the popular file. 
When there are not many devices which can deliver the video files successfully to users, it is better to focus on storing highly demanding videos.
Especially for SNR = 10dB in Fig. \ref{fig:CachingProb_vs_FileIndex_SNR}, files 4 and 5 will not be cached at any device.
Also, the caching probability gap between high-quality and low-quality files grows as $\lambda$ and SNR decrease, but the increments are not large, as long as the relative values of the quality measure to the file size is maintained.

\subsection{Queue Backlogs and Time-Average Quality Level with Optimal Node Association}

In this subsection, we examine the numerical results to verify the proposed node association algorithm.
All parameters of video files and storage size are the same as the prior subsection, but SNR=20 dB and $\lambda = 0.2$ are basically used here.
Additionally, we set $\mathcal{B}=1MHz$, $\tau_c = 5\times 10^{-3}$, $V=0.01$, $\tilde{Q}=10^2$, and $m=1$ so $\tau = \tau_c$.
Numerical results in this subsection are based on the system model of Fig. \ref{Fig:NodeAssociationCases}. 
$N=9$ users are assumed to be located in a grid structure as shown in Fig. \ref{Fig:NodeAssociationCases}, and the nearest users are separated by a distance of $d=20$.
If $R\leq d/2=10$, there will be no request collision, because coverage regions of users do not overlap. 
However, if $R \geq d/2=10$, request collision can occur.
For NOMA, the fixed power allocation ratios $\beta=[0.8, 0.2]$ and $\beta = [9/13, 3/13, 1/13]$ are assumed for the 2-user and 3-user cases, respectively.
Grouping more than three users for NOMA transmission is very rare, so NOMA for more than three users is not considered here.

To verify the advantages of the proposed node association algorithm, this paper compares the proposed one with two other comparison schemes:
\begin{itemize}
	\item `Maximum Arrival': The file-requesting user associates with the cache-enabled device which provides maximum arrivals within radius $R$.
	\item `Highest-Quality': The file-requesting user associates with the cache-enabled device which caches the requested file of the highest-quality, within radius $R$.
\end{itemize}

\begin{figure} [h!]
	\centering
	\includegraphics[width=0.4\textwidth]{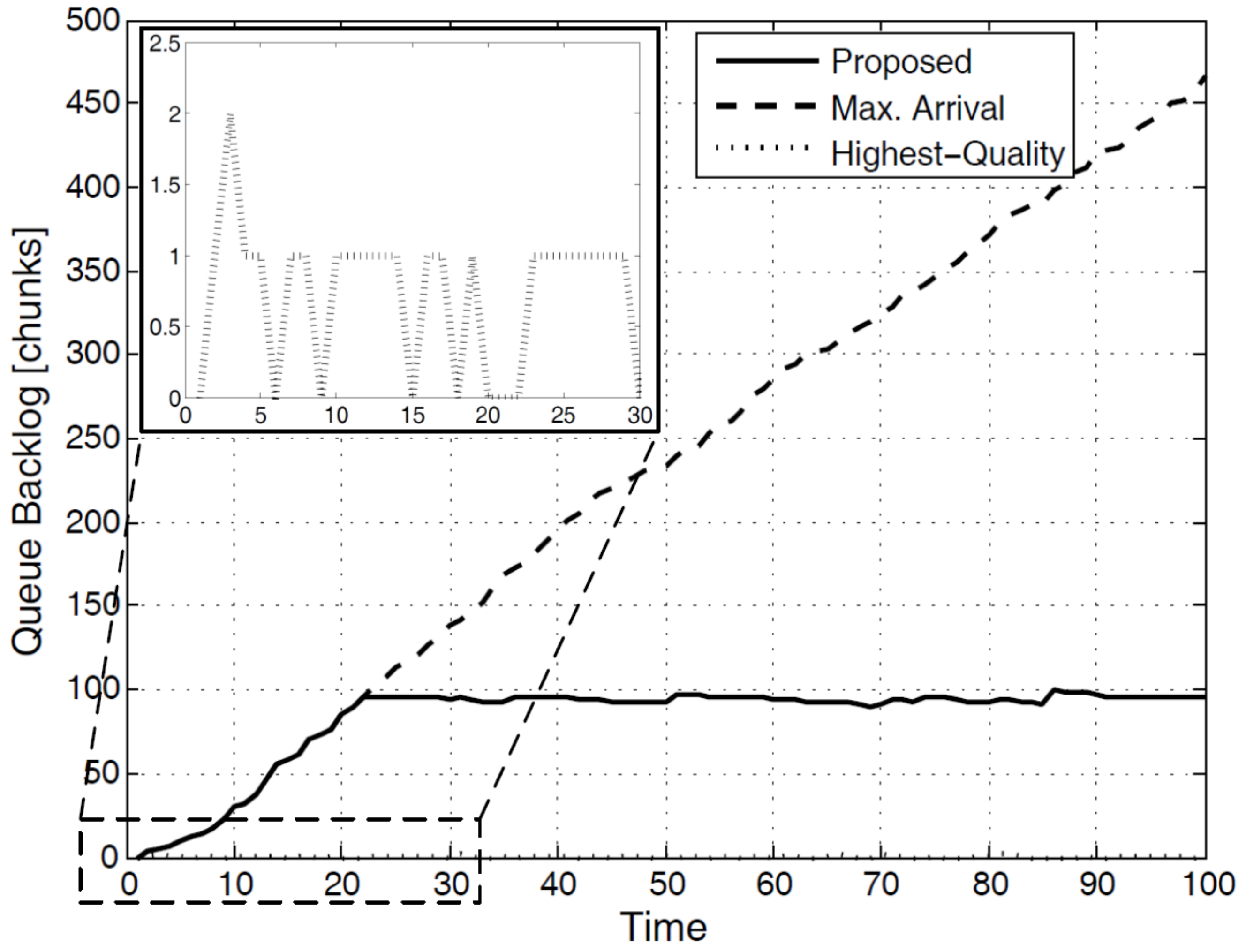}
	\caption{Queue backlog comparisons among node association schemes}
	\label{fig:backlog_comparison}
\end{figure}

Fig. \ref{fig:backlog_comparison} gives the plots of queue backlogs, i.e. the number of video chunks stacked in queue, versus time slot. 
The largest backlogs are stacked with the `Max-Arrival' scheme, the proposed algorithm is the next, and backlogs are hardly accumulated with the `Highest-Quality' scheme, as shown in Fig. \ref{fig:backlog_comparison}.
The `Max-Arrival' scheme communicates with the link providing the largest number of chunks, 
so it does not have to worry about playback delay, compared to other schemes.
For the proposed algorithm, the smaller video chunks are stacked in queue than `Max-Arrival', but its backlogs are large enough to avoid playback latency.
Specifically, Fig. \ref{fig:backlog_comparison} shows the effect of $\tilde{Q}$ which limits the maximal backlogs.
Since $\tilde{Q}=100$ chunks are enough to avoid playback delay, there is no need to stack too many chunks like `Max. Arrival'. 
Therefore, when $Q_n[t] \approx \tilde{Q}$, the proposed algorithm strongly pursues the high-quality file even though the D2D link of the device with the high-quality file is not good.
In addition, the queue size is finite in practical, the use of $\tilde{Q}$ can prevent queue overflow.
The `Highest-Quality' scheme has very little margin of the backlog for smooth video playback, and its enlarged graph is also shown in Fig. \ref{fig:backlog_comparison}.
The user queue of `Highest-Quality' scheme is frequently empty, thus several occurrences of playback delay are expected.

\begin{figure} [h!]
	\centering
	\includegraphics[width=0.37\textwidth]{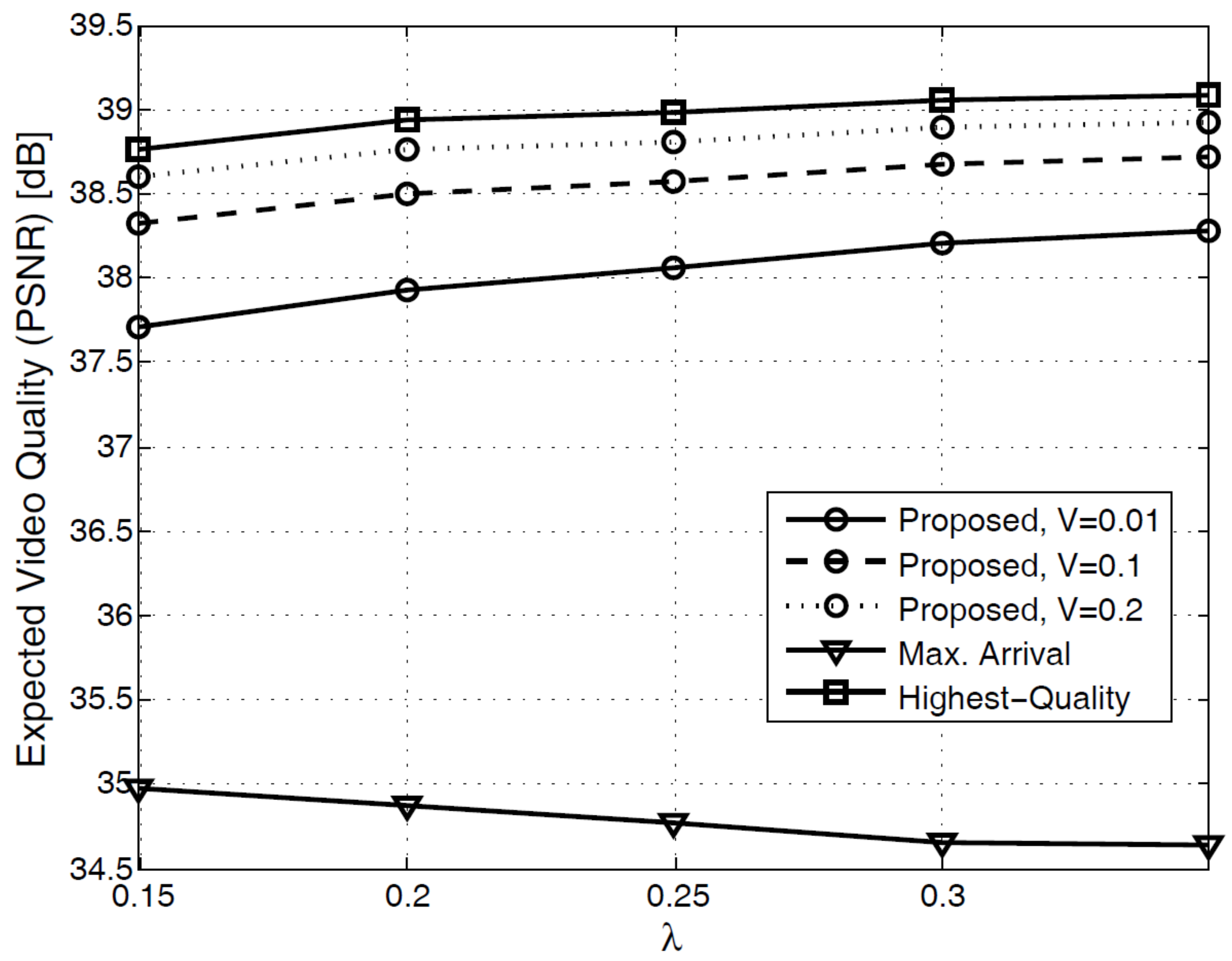}
	\caption{Time-averaged video quality measures with different values of $\lambda$}
	\label{fig:q_measure_lambda}
\end{figure}

\begin{figure} [h!]
	\centering
	\includegraphics[width=0.37\textwidth]{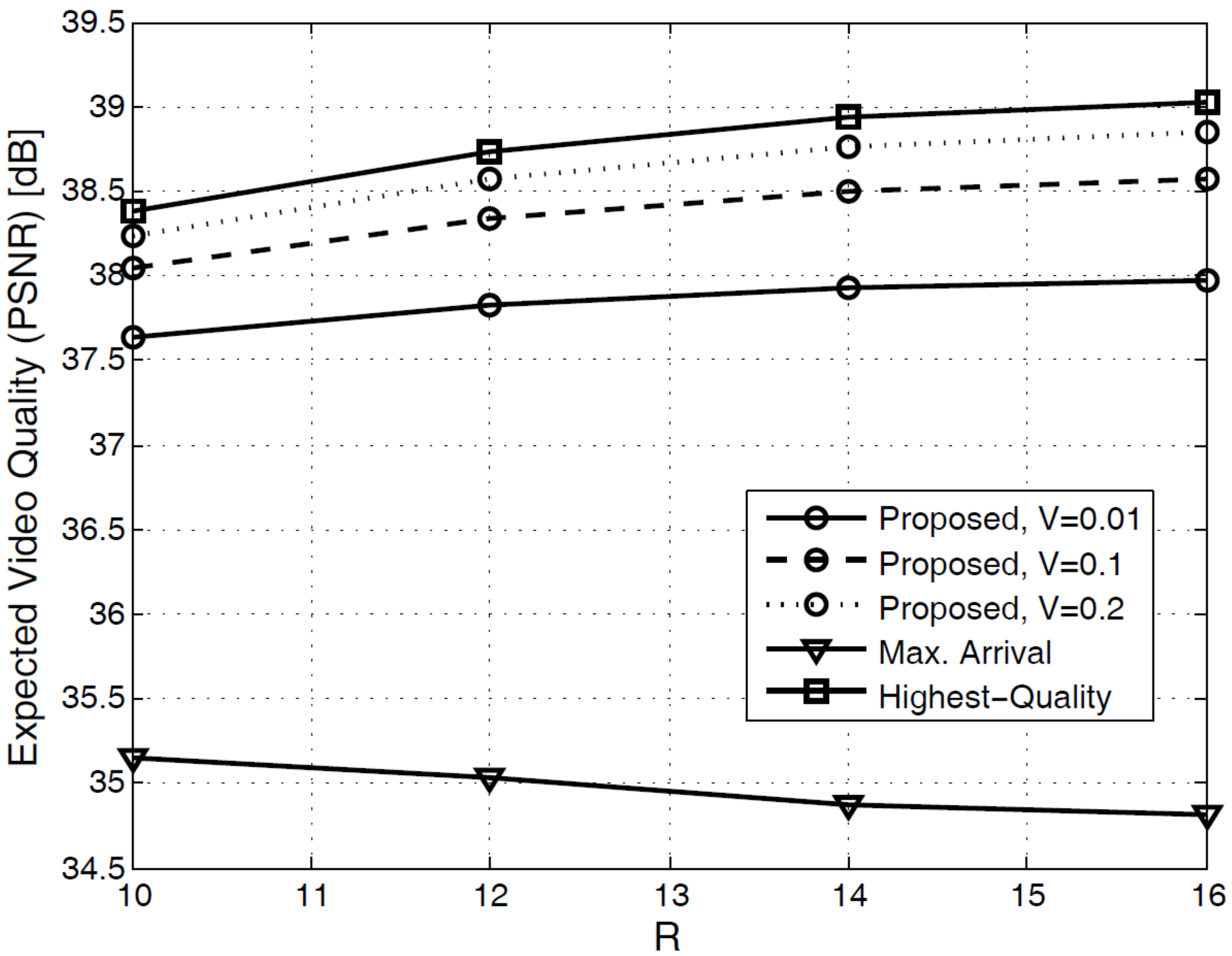}
	\caption{Time-averaged video quality measures with different values of $R$}
	\label{fig:q_measure_R}
\end{figure}

\begin{figure} [h!]
	\centering
	\includegraphics[width=0.37\textwidth]{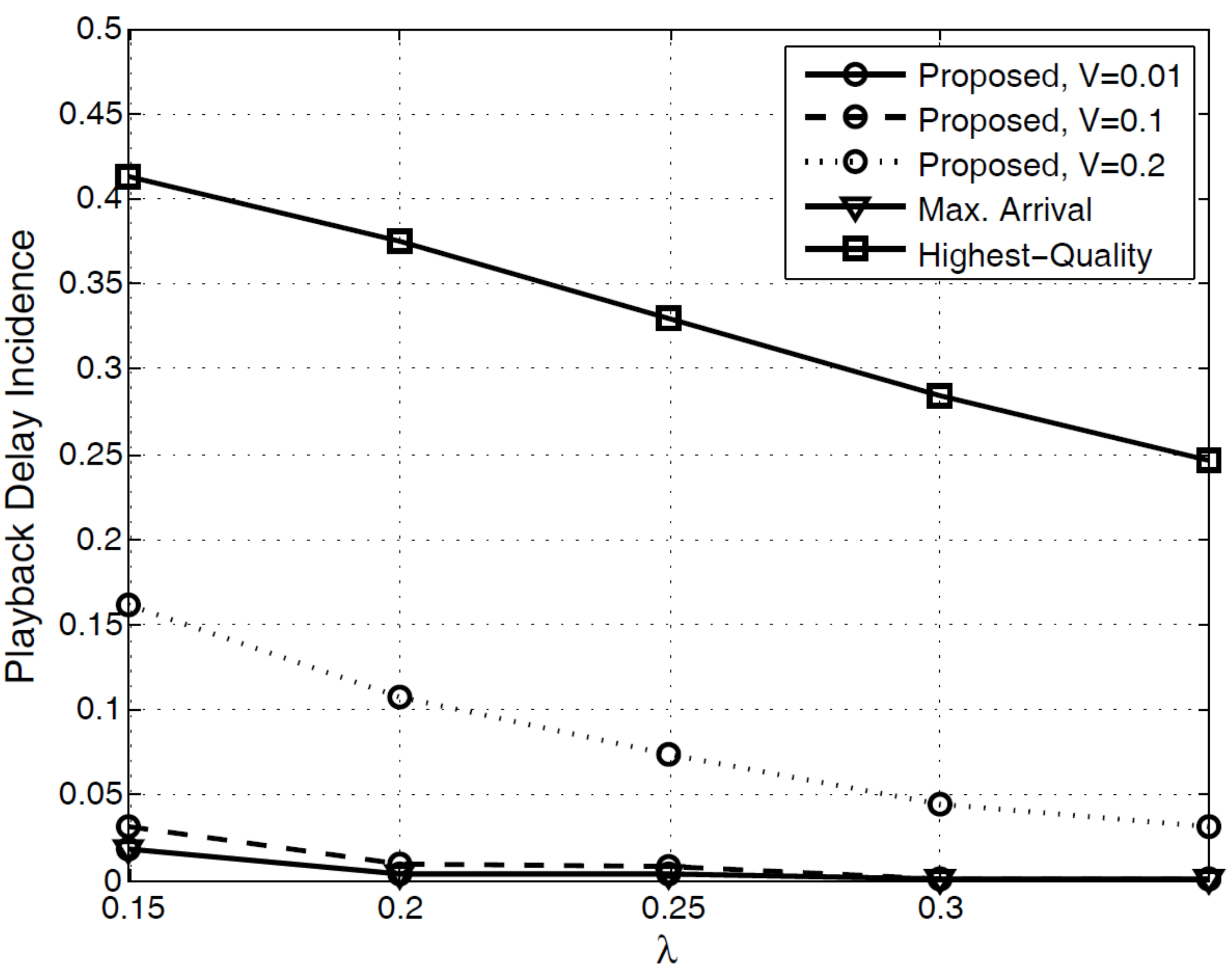}
	\caption{Playback delay incidence with different values of $\lambda$}
	\label{fig:delay_lambda}
\end{figure}

\begin{figure} [h!]
	\centering
	\includegraphics[width=0.37\textwidth]{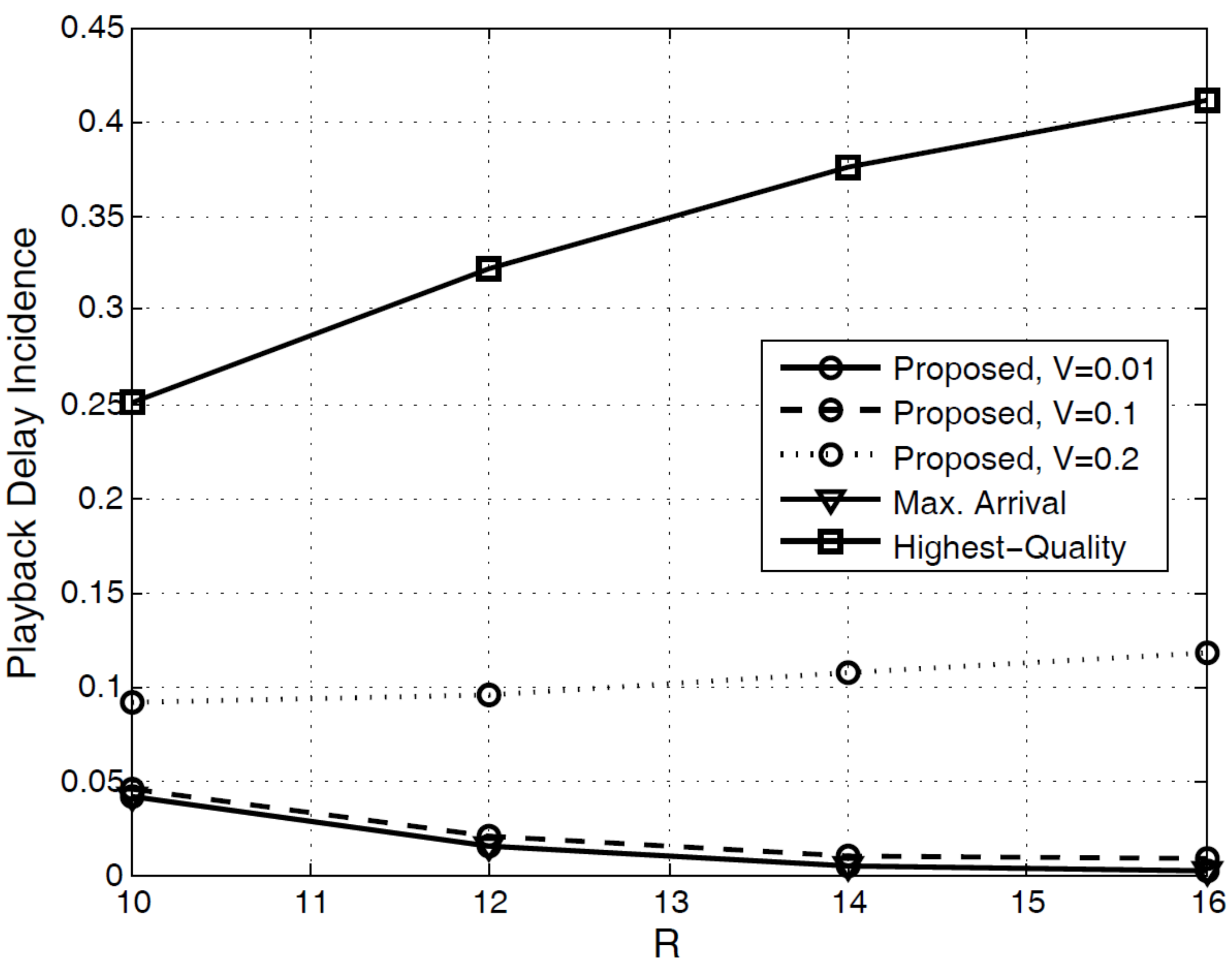}
	\caption{Playback delay incidence with different values of $R$}
	\label{fig:delay_R}
\end{figure}

The time-averaged video quality measures with different values of $\lambda$ and $R$ are shown in Figs. \ref{fig:q_measure_lambda} and \ref{fig:q_measure_R}, respectively. 
Obviously, the `Highest-Quality' scheme gives the best video quality.
Since the `Max. Arrival' scheme does not pursue video quality enhancement, it is obvious that its performance is the worst among the compared techniques in Figs. \ref{fig:q_measure_lambda} and \ref{fig:q_measure_R}.
The performance of the proposed algorithm is better than `Max-Arrival' and worse than `Highest-Quality'; however the proposed algorithm provides the similar quality measures to `Highest-Quality' as $V$ increases.
In addition, the performances of the proposed scheme and `Highest-Quality' improve with $\lambda$ or $R$ because these schemes pursue video quality enhancement.
On the other hand, `Max. Arrival' associates with the device which provides the maximal number of arrivals, preferring the strong channel and small file size (i.e., low-quality file). Thus, the quality measure of `Max. Arrival' degrades with $\lambda$ and $R$. 

Figs. \ref{fig:delay_lambda} and \ref{fig:delay_R} are plots of playback delay incidence versus $\lambda$ and $R$, respectively.
As we explained earlier, when there is no chunk in the queue while enjoying streaming service, playback delay occurs.
Therefore, playback delay incidence means how much queue emptiness occurs over the total playback time.
Among comparison schemes, the proposed algorithm with $V=0.01$ and `Max. Arrival' show the lowest playback delay, whereas there are much buffering times expected for the `Highest-Quality' scheme.
As $\lambda$ increases, more device candidates which can provide the desired file with good channel conditions are expected; thus delay incidences of all schemes decrease, whereas the trends of delay incidences in accordance with $R$ are different. 
In the `Highest-Quality' scheme, when $R$ is large and $\lambda$ is fixed, the distance between the streaming user and the device storing the best-quality file would be large, i.e., the associated device would experience a bad channel.
On the contrary, for the `Max. Arrival' scheme it becomes easier to find the device candidate which can deliver more chunks, assuming $R$ is large. 
Thus, delay incidence of `Highest-Quality' increases with $R$ whereas that of `Max. Arrival' decreases. 
In the case of the proposed algorithm, as $V$ becomes larger, the streaming user pursues the video quality rather than reduced playback delay. 
Thus, the delay incidence with $V = 0.2$ increases with $R$, whereas those with $V = 0.1$ and $V = 0.01$ do not. 

Considering the results of both quality measure and playback delay incidence, we can say that the proposed algorithm smooths out the trade-off between video quality and playback delay. 
The `Max-Arrival' scheme is good to avoid playback latency, but the file-requesting users would be suffered from the degraded video quality.
On the other hand, the `Highest-Quality' scheme provides the best video quality, but its user experiences too much buffering times to enjoy the smooth streaming service.
Thus, the proposed node association algorithm can be useful for achieving both acceptable playback delay and high enough video quality.
In addition, the trade-off between video quality and reduced playback delay in our proposed algorithm can be controlled by adjusting the system parameter $V$.
In Figs.~\ref{fig:q_measure_lambda} and \ref{fig:q_measure_R}, the expected video quality increases as $V$ grows up, whereas the playback delays occur more frequently as shown in Figs.~\ref{fig:delay_lambda} and \ref{fig:delay_R}.

\section{Concluding Remarks}
\label{sec:conclusion}

This paper considered video files of various quality levels in the D2D-assisted wireless caching network.
This paper suggests the optimal caching policy for video files of different quality levels and thus of different sizes which maximizes the successfully enjoyable quality sum. 
In addition, a node association algorithm has been proposed that maximizes the sum of the time-average video quality that file-requesting users enjoy while preventing playback delay, the most important user QoS in video streaming service. 
In this paper, \textit{request collision}, the situation where a device receives file requests from multiple users at the same time, has been considered, and the solutions based on NOMA as well as scheduling of one user have been presented. 
The proposed file placement rule and the node association algorithm have been verified by simulation results.



%

\begin{IEEEbiography}[{\includegraphics[width=1in,height=1.25in,clip,keepaspectratio]{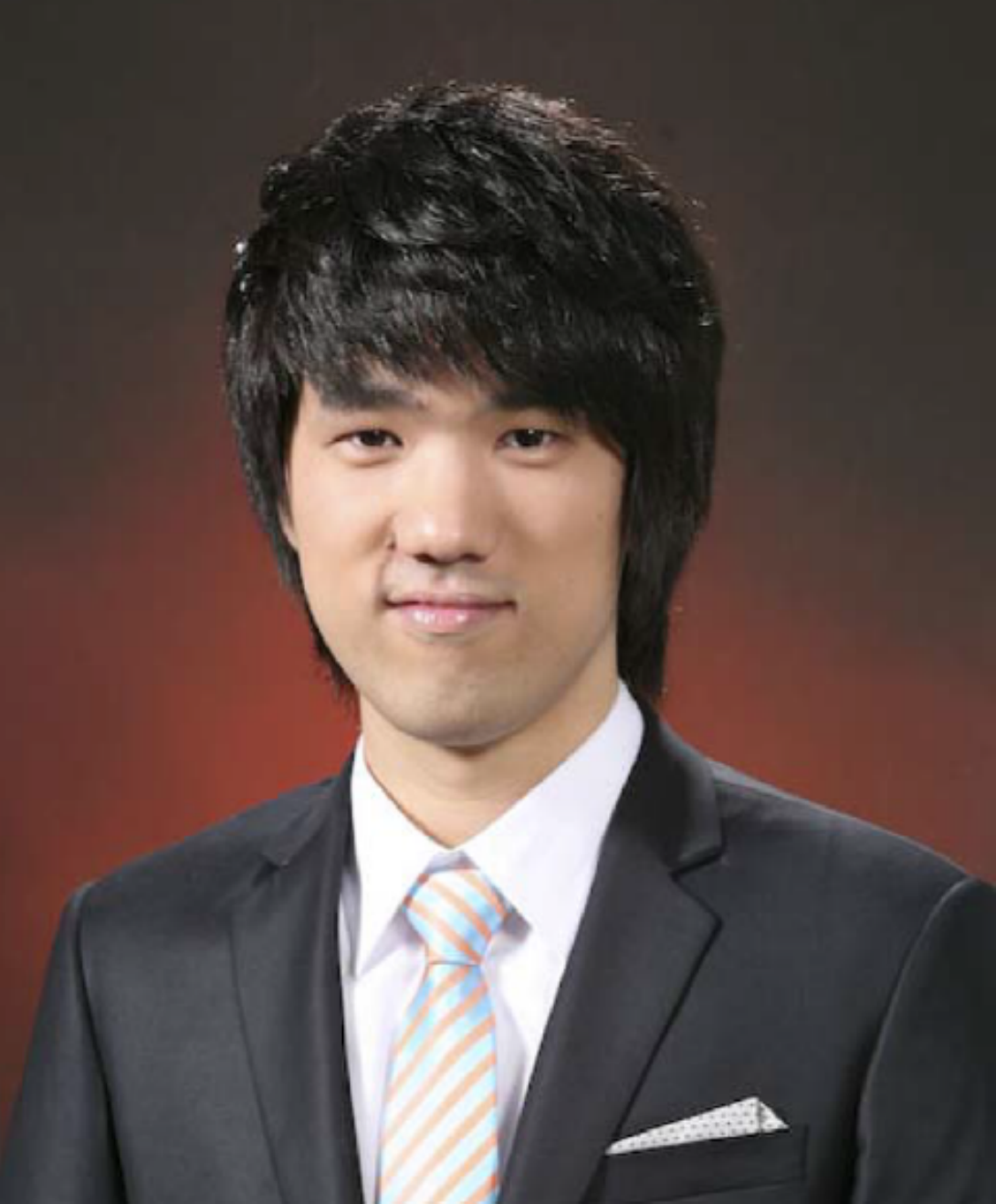}}]{Minseok Choi}
	received the B.S. and M.S. degree in electrical engineering from the Korea Advanced Institute of Science and Technology (KAIST), Daejeon, Korea, in 2016. He is currently pursuing the Ph.D degree in KAIST. His research interests include NOMA, Wireless caching network, 5G Communications, and mmWave.\end{IEEEbiography}
\begin{IEEEbiography}[{\includegraphics[width=1in,height=1.25in,clip,keepaspectratio]{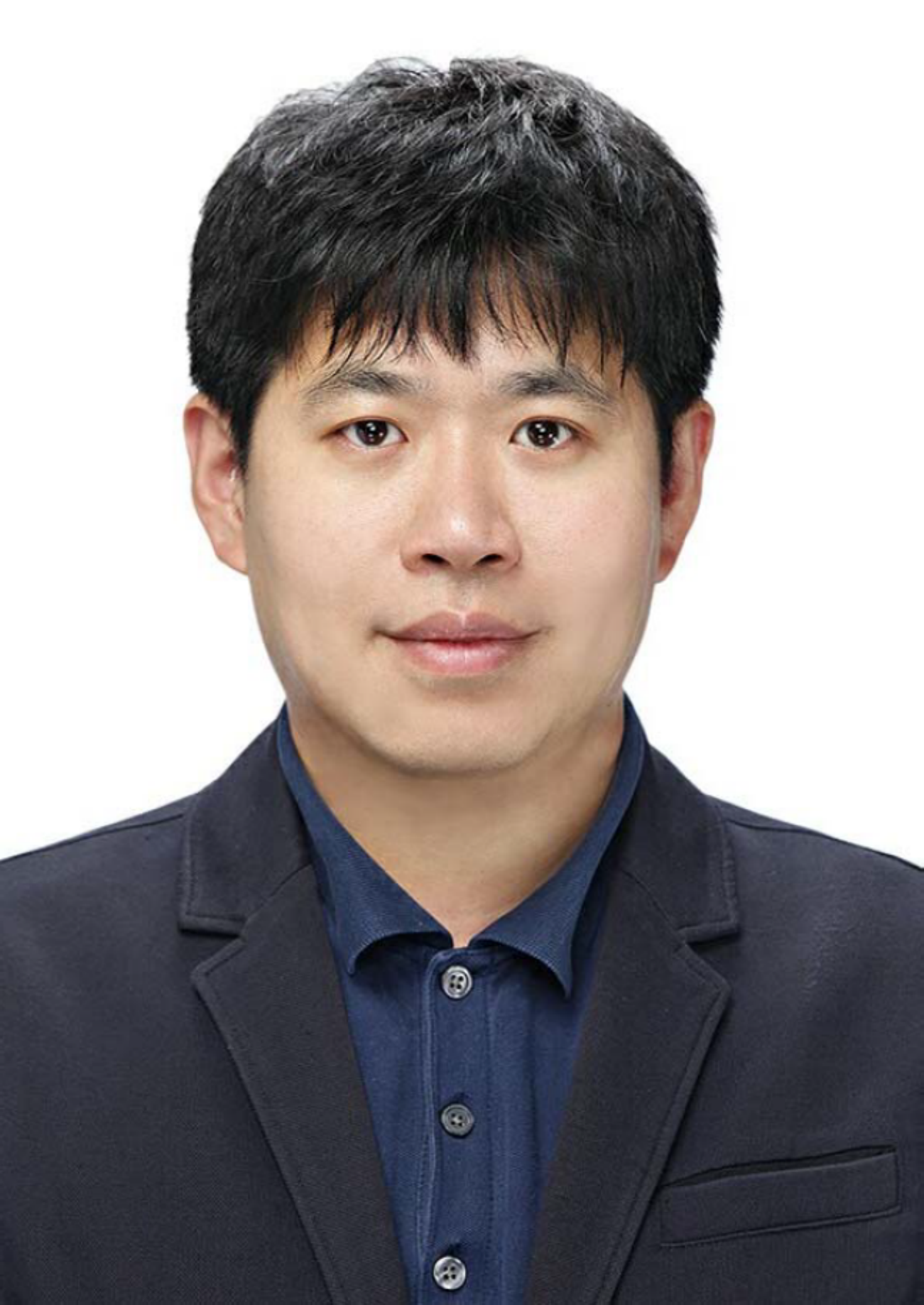}}]{Joongheon Kim}
	(M'06--SM'18) has been an assistant professor with Chung-Ang University, Seoul, Korea, since 2016. He received his B.S. (2004) and M.S. (2006) in computer science and engineering from Korea University, Seoul, Korea; and his Ph.D. (2014) in computer science from the University of Southern California (USC), Los Angeles, CA, USA. In industry, he was with LG Electronics Seocho R\&D Campus (Seoul, Korea, 2006--2009), InterDigital (San Diego, CA, USA, 2012), and Intel Corporation (Santa Clara, CA, USA, 2013--2016). 
	
	He is a senior member of the IEEE; and a member of IEEE Communications Society. He was awarded Annenberg Graduate Fellowship with his Ph.D. admission from USC (2009).\end{IEEEbiography}
\begin{IEEEbiography}[{\includegraphics[width=1in,height=1.25in,clip,keepaspectratio]{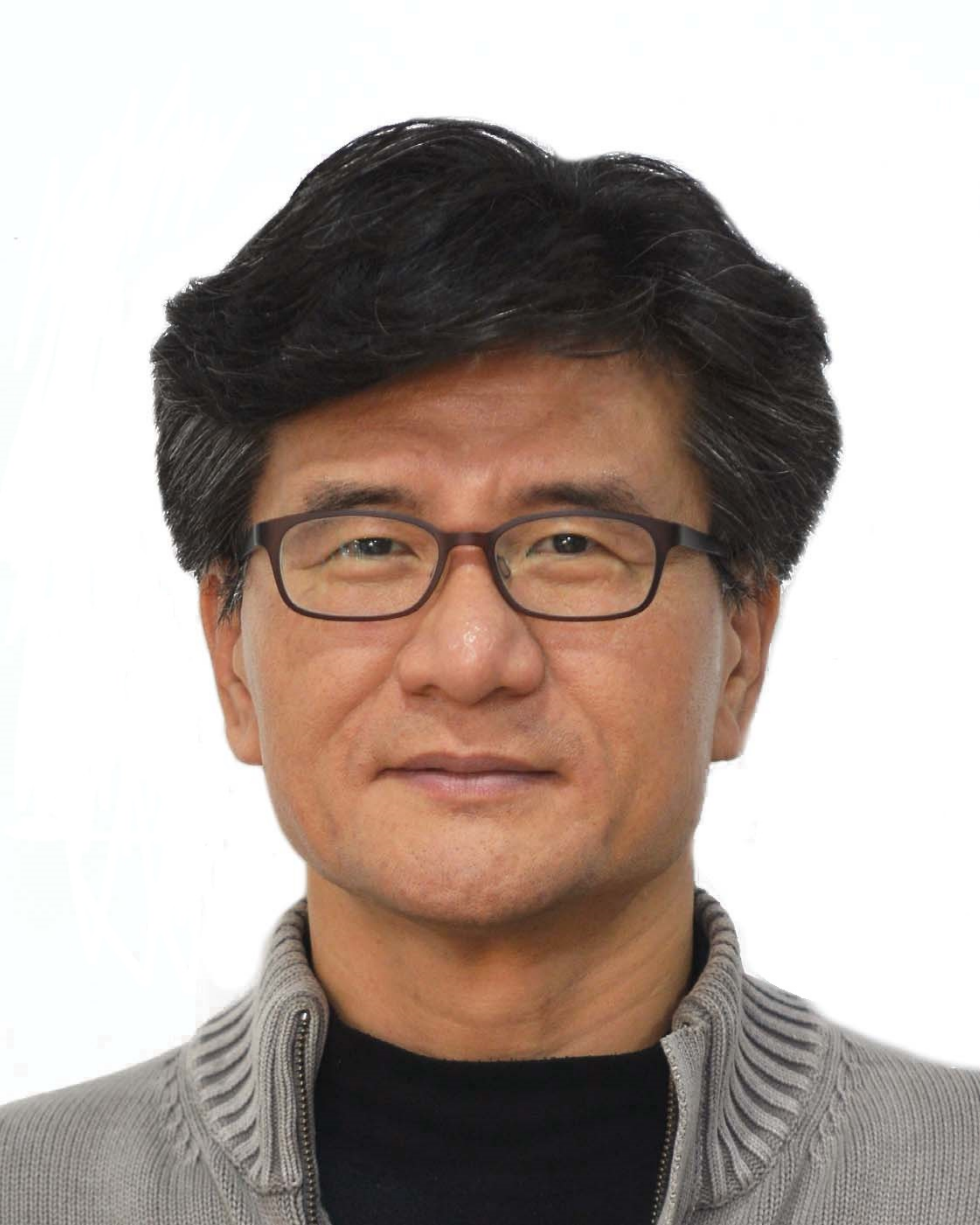}}]{Jaekyun Moon}
	received the Ph.D degree in electrical and computer engineering at Carnegie Mellon University, Pittsburgh, Pa, USA. He is currently a Professor of electrical enegineering at KAIST. From 1990 through early 2009, he was with the faculty of the Department of Electrical and Computer Engineering at the University of Minnesota, Twin Cities. He consulted as Chief Scientist for DSPG, Inc. from 2004 to 2007. He also worked as Chief Technology Officier at Link-A-Media Devices Corporation. His research interests are in the area of channel characterization, signal processing and coding for data storage and digital communication. Prof. Moon received the McKnight Land-Grant Professorship from the University of Minnesota. He received the IBM Faculty Development Awards as well as the IBM Partnership Awards. He was awarded the National Storage Industry Consortium (NSIC) Technical Achievement Award for the invention of the maximum transition run (MTR) code, a widely used error-control/modulation code in commercial storage systems. He served as Program Chair for the 1997 IEEE Magnetic Recording Conference. He is also Past Chair of the Signal Processing for Storage Technical Committee of the IEEE Communications Society. He served as a guest editor for the 2001 IEEE JSAC issue on Signal Processing for High Density Recording. He also served as an Editor for IEEE TRANSACTIONS ON MAGNETICS in the area of signal processing and coding for 2001-2006. He is an IEEE Fellow.
\end{IEEEbiography}\vfill




\end{document}